\documentclass{article}

\usepackage{graphicx}
\usepackage{url}
\usepackage{amsmath}
\usepackage[all]{nowidow}

\usepackage{subcaption}

\usepackage{float}
\usepackage{authblk}

\begin{document}

\title{AuctionWhisk: Using an Auction-Inspired Approach for Function Placement in Serverless Fog Platforms}

\author[1]{David Bermbach}

\author[2]{Jonathan Bader}

\author[1]{Jonathan Hasenburg}

\author[1]{Tobias Pfandzelter}

\author[2]{Lauritz Thamsen}

\affil[1]{Mobile Cloud Computing Research Group\\TU Berlin \& Einstein Center Digital Future\\Berlin, Germany\\\texttt{\{db,jh,tp\}@mcc.tu-berlin.de}}

\affil[2]{Distributed and Operating Systems Research Group\\TU Berlin\\\texttt{\{jonathan.bader,lauritz.thamsen\}@tu-berlin.de}}

\date{}

\maketitle

\abstract{
    The Function-as-a-Service (FaaS) paradigm has a lot of potential as a computing model for fog environments comprising both cloud and edge nodes, as compute requests can be scheduled across the entire fog continuum in a fine-grained manner.
    When the request rate exceeds capacity limits at the resource-constrained edge, some functions need to be offloaded towards the cloud.

    In this paper, we present an auction-inspired approach in which application developers bid on resources while fog nodes decide locally which functions to execute and which to offload in order to maximize revenue.
    Unlike many current approaches to function placement in the fog, our approach can work in an online and decentralized manner.
    We also present our proof-of-concept prototype AuctionWhisk that illustrates how such an approach can be implemented in a real FaaS platform.
    Through a number of simulation runs and system experiments, we show that revenue for overloaded nodes can be maximized without dropping function requests.
}

\section{Introduction}\label{sec:intro}
Recently, the paradigm of fog computing has received more and more attention.
In fog computing, cloud resources are combined with resources at the edge, i.e., near the end user or close to IoT devices, and in some cases also with additional resources in the network between cloud and edge~\cite{paper_bermbach_fog_computing}.
While this adds complexity for applications, it comes with three key benefits: First, leveraging compute resources at or near the edge can lower response times, which is crucial for application domains such as autonomous driving or 5G mobile networks~\cite{paper_bermbach_fog_computing,BonomiFog}.
Second, data can be filtered and pre-processed early on the path from edge to the cloud, which reduces the data volume~\cite{lu2017lightweight}.
Especially in IoT use cases, it is often not feasible to transmit all data to the cloud as the sheer volume of produced data exceeds the bandwidth capabilities of the network~\cite{paper_zhang_cloud_is_not_enough_GDP} or leads to significant energy consumption for wide-area networking~\cite{wiesner2021leaf}.
Third, keeping parts of applications and data at the edge can help to improve privacy, e.g., by avoiding ``centralized data lakes''~\cite{paper_pallas_fog4privacy} in the cloud.
Overall, fog computing, thus, combines the benefits of both cloud and edge computing.

While there are many open research questions in fog computing, a key question has not been answered yet:
Which compute paradigms will future fog applications follow?
In previous work~\cite{paper_bermbach_fog_computing}, we argued that a serverless approach -- which we understand as Function-as-a-Service (FaaS) in this paper -- is a good fit for the edge.
The main reason for this is that resources at the edge are regurlarly considerably constrained so that provisioning them in small function slices is more efficient than provisioning them using virtual machines or long-running containers.
Additionally, the idea of having strictly stateless functions, separated from data management~\cite{paper_baldini_openwhisk,paper_hellerstein_serverless}, supports moving parts of applications seamlessly between edge and cloud resources.

Now, assuming a serverless world in which application components can run as functions on FaaS platforms in the cloud, at the edge, and medium-sized data centers in between, the question of how to distribute fog application components can be reduced to the issue of function placement across multiple geo-distributed sites.
In~\cite{paper_bermbach_auctions4function_placement}, we introduced the idea of using auction-inspired mechanisms for function placement and evaluated it in small-scale simulation experiments.
We were able to confirm our assumptions that this would be an efficient approach to decentralized scheduling of functions over a distributed fog infrastructure and have argued for further work in this area.
In this paper, we extend our previous work and make the following contributions:

\begin{enumerate}
    \item We describe a more general conceptual approach of using a decentralized auction scheme to control function placement (Section~\ref{sec:approach}).
    \item We discuss practical engineering challenges for implementing this approach in existing FaaS platforms (Section~\ref{sec:challenges})
    \item We present a simulation tool as well as the underlying system model and use it to study the effects of various parameters on our auction-based function placement (Section~\ref{sec:eval_sim}).
    \item We implement our approach as a proof-of-concept prototype called AuctionWhisk based on Apache OpenWhisk and evaluate our prototype through experiments (Section~\ref{sec:eval_openWhisk}).
    \item We critically discuss the limitations of our work and identify future research directions in the field (Section~\ref{sec:discussion}).
\end{enumerate}

\section{Background}\label{sec:background}
In this section, we introduce and describe fundamental concepts of FaaS, fog computing, and auctions.
As all three topics have received major attention in research in the last few years and different definitions have emerged, we want to clarify the terminology we adopt in this paper.
We also give a short overview of Apache OpenWhisk which we used as basis for our prototype system.

\subsection{Function-as-a-Service (FaaS)} \label{subsec:faas}
Within the field of cloud computing, the Function-as-a-Service (FaaS) paradigm has emerged as the latest evolution in resource sharing.
The FaaS programming model facilitates highly scalable event-driven applications.

Developers deploy their code in the form of functions to a FaaS platform that handles code invocation and scaling, lowering the management burden for the consumer.
Here, a function is a piece of business logic that is executed in response to an event.
Functions can be implemented in any programming language as long as a runtime environment for the language is supported by the target FaaS platform.
Events can be web requests, monitoring data, or even IoT sensor readings, thus making the FaaS approach a versatile option for many use cases.
Logically, these functions live only as long as they process a single event and are reset with every invocation.
This is usually achieved with lightweight virtualization techniques such as containerization, microVMs, or unikernels~\cite{paper_pfandzelter_tinyfaas,Agache2020-ug,paper_pfandzelter_streams_functions}, which enable FaaS platforms to spin up and destroy isolated instances quickly.
As a result, no state can be persisted within a function across multiple invocations, they are thus referred to as ``stateless''.
This is also one of the main reasons for their scalability:
A FaaS platform can spin up multiple concurrent instances of a function to handle concurrent events, while quickly shutting them down to free up resources for other functions.
The same characteristic makes FaaS functions a good fit to fog and edge environments~\cite{paper_bermbach_fog_computing}:
When migrating functions, there is no need to handle session state within the function container.
Instead, the function container can simply be terminated on the original node and restarted on the new node.
In practice, FaaS functions are usually used in conjunction with other platform services, especially in combination with storage and database services that manage state, e.g.,~\cite{paper_confais_ipfs4fog}.

While scalability and low management overhead are clear advantages for developers, operators have the benefit that their infrastructure can be leveraged more efficiently.
Instead of allocating coarsely grained resources in the form of virtual machines or containers statically, these more finely grained functions can be more efficiently mapped to underlying infrastructure and moved dynamically~\cite{Baldini2017-uu,paper_hendrickson_openlambda,Lynn2017-bz}.
The platform provider can then charge tenants based on actual usage.
In 2021, all major cloud service providers offer such a FaaS platform, e.g., AWS Lambda\footnote{aws.amazon.com/lambda}, Microsoft Azure Functions\footnote{azure.com/functions}, Google Cloud Functions\footnote{cloud.google.com/functions}, or IBM Cloud Functions\footnote{www.ibm.com/cloud/functions}.

\subsection{Apache OpenWhisk}
\label{subsec:apache_openWhisk}

In addition to cloud-hosted FaaS platforms, a number of options for FaaS platforms have also emerged in research and open source communities, e.g., tinyFaaS~\cite{paper_pfandzelter_tinyfaas}, OpenLambda~\cite{paper_hendrickson_openlambda}, or SAND~\cite{paper_akkus_sand}.
Another noteworthy example is Apache OpenWhisk, which is also at the core of IBM's Cloud Function service~\cite{paper_baldini_openwhisk}.

\begin{figure}
    \centering
    \includegraphics[width=0.7\columnwidth]{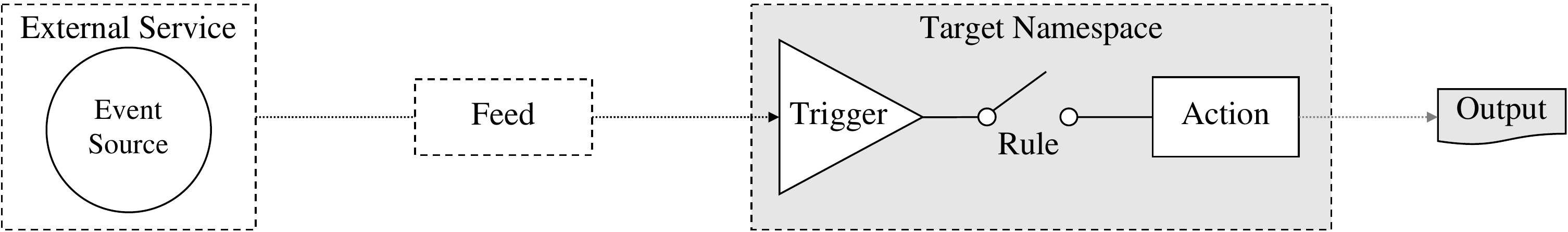}
    \caption{The OpenWhisk programming model~\cite{The_Apache_Software_Foundation_undated-ew}.}
    \label{fig:ow_programming}
\end{figure}

The OpenWhisk platform allows for custom functions, called \emph{actions}, to be executed in response to an event, the \emph{trigger}.
Figure~\ref{fig:ow_programming} gives a high-level overview of the programming model as shown in the official OpenWhisk documentation.

\begin{figure}
    \centering
    \includegraphics[width=0.7\columnwidth]{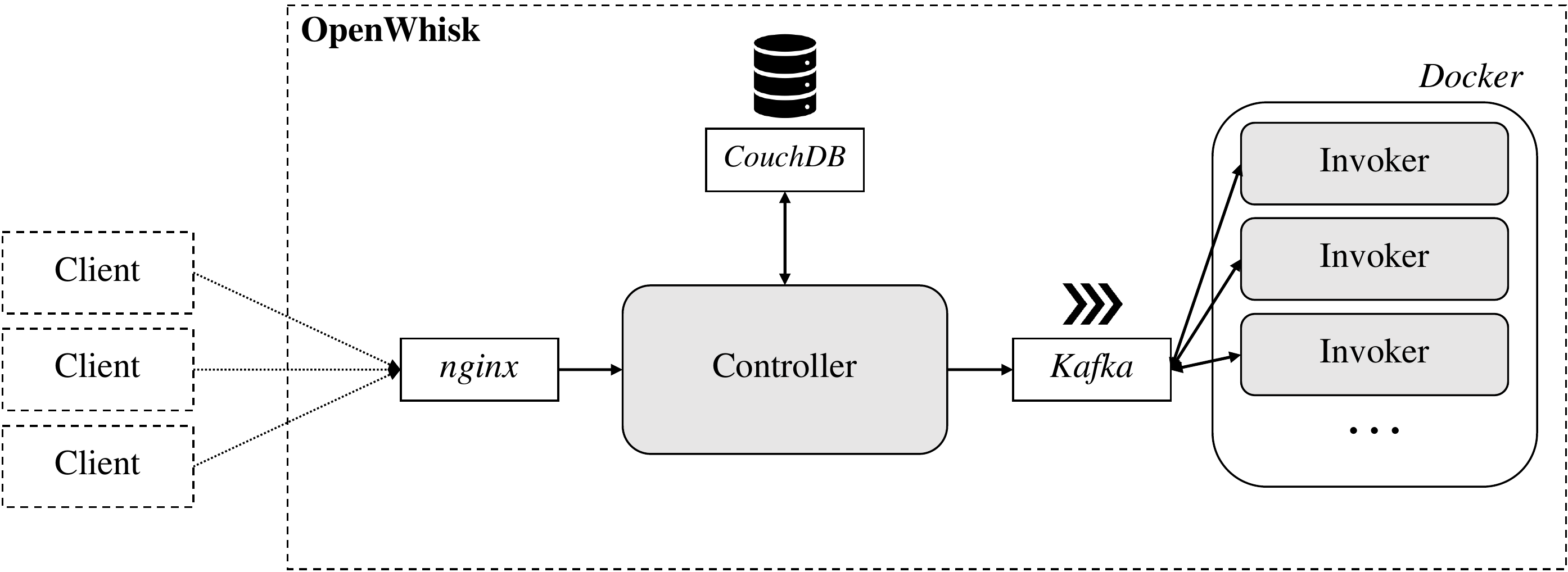}
    \caption{Components of an OpenWhisk deployment~\cite{The_Apache_Software_Foundation_undated-ew}.}
    \label{fig:ow_deployment}
\end{figure}

Similarly, Figure~\ref{fig:ow_deployment} gives an overview of the OpenWhisk components:
At the core of the OpenWhisk platform, the \emph{Controller} provides endpoints to process triggers and schedules action invocations.
The \emph{Invokers} use Docker containers to isolate different actions, with a new container being spawned for each execution request.
A \emph{Kafka} queue serves as a buffer between Controller and Invoker, e.g., to avoid unnecessary cold starts~\cite{paper_bermbach_faas_coldstarts}.
Additionally, a \emph{CouchDB} instance holds actions, triggers, rules, and user-related information such as credentials or namespaces; \emph{nginx} is used as the HTTP endpoint for the platform.

A complete installation of the OpenWhisk platform requires a substantial amount of resources which may not always be available to the FaaS platform, especially when moving towards the edge.
For this reason, there is also a more lightweight version of OpenWhisk -- \emph{Lean} OpenWhisk.
Lean OpenWhisk removes components such as the CouchDB database and Kafka queue in order to leave more resources for the function instances at the cost of some platform features.

\subsection{Fog Computing}
\begin{figure}
    \centering
    \includegraphics[width=0.7\columnwidth]{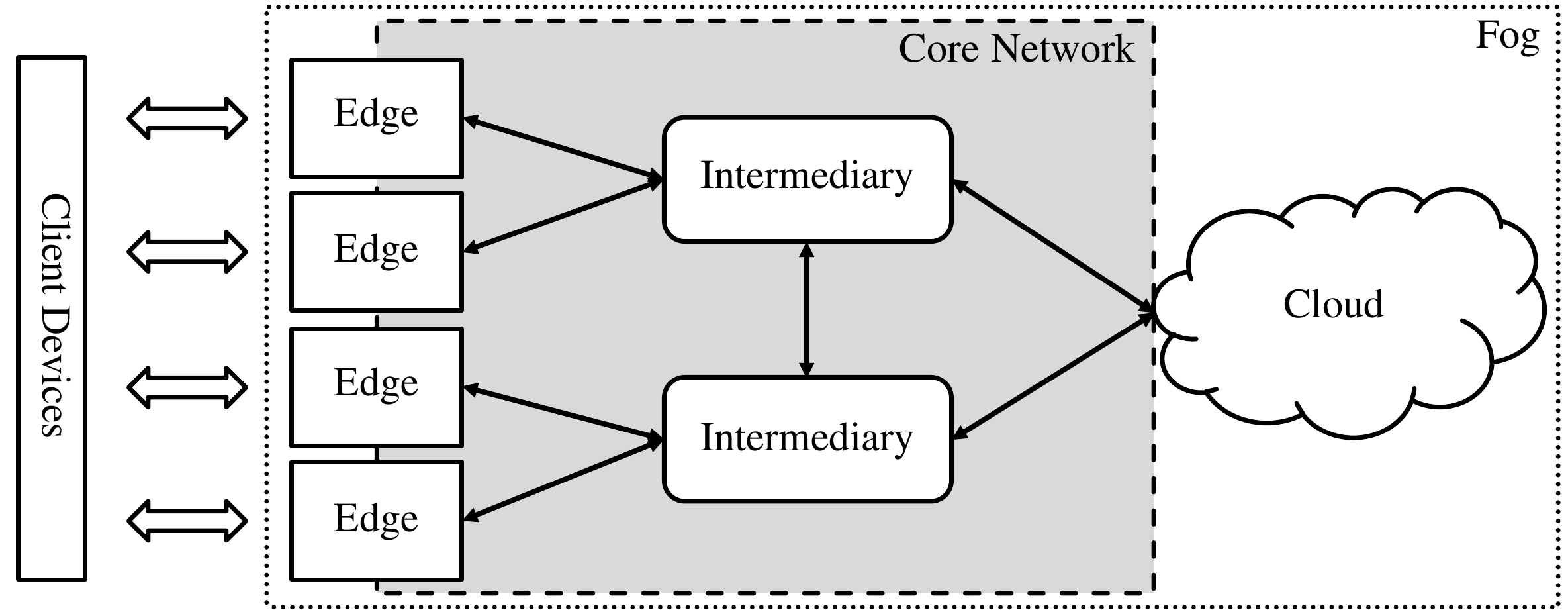}
    \caption{Overview of Cloud, Edge, and Fog Computing (adapted from~\cite{paper_bermbach_fog_computing})}
    \label{fig:fog}
\end{figure}

As the interest in fog computing increased, we have seen the emergence of different, often conflicting definitions of the term.
In some cases, it is used as a synonym for edge computing, in others it is used to refer to resources that sit between edge and cloud or to all resources between cloud and device.
In this paper, we adopt the definition of~\cite{paper_bermbach_fog_computing}, where fog resources encompass all resources in edge, cloud, and in between, yet not those of end devices.
We refer to these end devices also as ``clients'' and mean (embedded) IoT devices as well as mobile phones or computers, and to resources between edge and cloud as ``intermediary nodes'' or ``intermediary'' in short.
We show an overview of cloud, edge, and fog computing in Figure~\ref{fig:fog}.

This definition yields transparency from a client perspective, as clients simply access fog resources unaware of whether a particular service resides at the edge or in the cloud.
Rather, they only see a singular ``fog''.

In reality, the implementation of the fog is expected to follow a hierarchical model, where small edge nodes are collocated with access network equipment such as radio towers and fewer yet larger intermediary nodes are placed within the core network.
Edge nodes will usually have the capabilities of single hosts such as Raspberry Pis or a small local cluster, while intermediaries are small- to medium-sized data centers.
At the root of the tree-like model is the cloud with its seemingly infinite compute and storage resources.
This leads to tradeoff decisions that have to be made for each application:
on the one end, edge nodes can provide low latency, high bandwidth resource access for client devices, albeit with less available or more expensive resources.
At the other end, the cloud provides scalable and (seemingly) infinite resources, but at the cost of large network distances to clients.
Intermediary fog nodes fall somewhere in between.

\subsection{Fog-Based FaaS Platforms}

We expect the paradigm of fog computing to enable entirely new classes of services and also to increase the quality-of-service (QoS)~\cite{book_cloud_service_benchmarking} of existing applications.
In order to leverage the capabilities of the fog, it has been proposed to deploy applications on FaaS platforms that run on the fog infrastructure~\cite{paper_pfandzelter_streams_functions,paper_pfandzelter_tinyfaas,paper_baresi_serverless4fog,paper_george_nanolambda,hall2019execution,aslanpour2021serverless,paper_pfandzelter_LEO_serverless}.

Open source platforms such as OpenWhisk or OpenLambda~\cite{paper_hendrickson_openlambda} are a good fit for cloud nodes and larger intermediary nodes, yet their larger deployment overheads make them unsuitable for smaller intermediaries or the edge~\cite{palade2019evaluation}.
On more constrained nodes, Lean OpenWhisk and tinyFaaS~\cite{paper_pfandzelter_tinyfaas} are possible options for function deployment, whereas NanoLambda is an option for on-device FaaS deployment~\cite{paper_george_nanolambda}.

\subsection{Auctions\label{sec:auctions}}
When demand for a product or service exceeds its available supply, auctions can be used to reach market equilibrium (i.e., to match suppliers with the optimal buyers).
In an auction, all potential buyers enter bids which ideally correspond to the amount they are willing to pay.
An auctioneer then sets auction rules which determine (i) how the winning bid(s) are chosen and (ii) which price the successful buyers have to pay.

There is a large body of research on different auction rules.
Here, we focus on the so-called sealed-bid auctions in which potential buyers enter their bids hidden from other potential buyers and the winner (or winners if there is more than one auctioned item) of the auction is the buyer with the highest bid.
In these sealed-bid auctions, there are two fundamental rules for setting the price: in first-price auctions, the winner pays its bid; in second-price auctions, the winner pays the bid of the second highest (unsuccessful) potential buyer.
The latter strategy is recommended whenever it is important that bids are truthful, i.e., they correspond to the price that the potential buyers consider fair.

In the rest of this paper, we will not expressly differentiate between different auction types but will implicitly assume either a first-price or second-price sealed auction.
In the approach that we will present, it does not matter which pricing rule is used -- they may affect prices but will not affect whether a placement decision is reached.

\section{Auction-Based Function Placement}\label{sec:approach}
While a number of FaaS systems is already available, as discussed in Section~\ref{subsec:faas}, it is still not clear how to connect different FaaS deployments.
Ideally, we would want edge, intermediary, and cloud deployments to coordinate function placement among each other.
For instance, a request arriving in the cloud should probably still be processed in the cloud.
A request arriving at the edge, however, should in most cases (exceptions include functions that require data located elsewhere~\cite{paper_hellerstein_serverless}) be executed right at the edge.
Only when the load on the edge node exceeds the available capacity should the request be delegated to an intermediary (and likewise from intermediaries towards the cloud).
This concept, similar to cloud bursting, is based on the intuition of the cloud practically providing infinite resources~\cite{nist_definition_cloud_computing}.

Based on this, we can conclude that function placement is straightforward when nodes have spare capacity (see also~\cite{paper_pfandzelter_streams_functions} for a more general discussion) but becomes challenging when, for instance, an edge node is overloaded.
In such a situation, the question is which request should be delegated to the next node on the path to the cloud as that request will incur extra latency.
For reasons of resilience and fault-tolerance in a geo-distributed fog environment, such scheduling should ideally be managed in a decentralized manner, i.e., through local decisions on each node.
We can imagine a number of objectives and criteria for such local decisions, e.g., to prefer short-running functions over long-running ones (and vice versa), to consider bandwidth impacts, to give some clients preference over others, or to prefer latency-critical functions over less critical ones.

In contrast to these, we propose to use an auction-inspired approach, which has been shown to lead to an efficient resource allocation in multiple domains (e.g.,~\cite{Huang08:ABR,Gao11:MAP,Zhao14:HCT,Yang16:IMC,Kantarci14:TSP,Zhang14:DRP,Chandrashekar07:ABM,Jayaweera11:ACC}):
When application developers deploy their function to an integrated fog FaaS platform, they also attach two bids to the executable.
The first bid is the price that the respective developer is willing to pay for a node to store the executable (in \$/s), the second bid is for the actual execution of the function (in \$/execution).
In practice, both bids are likely to be vectors so that developers could indicate their willingness to pay more on edge nodes than in the cloud or even on a specific edge node.
We explicitly distinguish bids for storage of the executable and the execution as edge nodes might encounter storage limits independently from processing limits. Both auctions closely resemble a sequence of first-price sealed-bid auctions as discussed in Section~\ref{sec:auctions}.

For ease of explanation, we use a first-price auction, although a second-price auction would be preferable in practice.
Furthermore, as there is likely to be a minimum price equivalent to today's cloud prices, we can consider both bids a surcharge on top of a constant regular cloud price.
For sake of clarity and since it has no impact on the bids, we leave this constant price aside in our explanation.

\begin{figure}
    \centering
    \includegraphics[width=0.6\columnwidth]{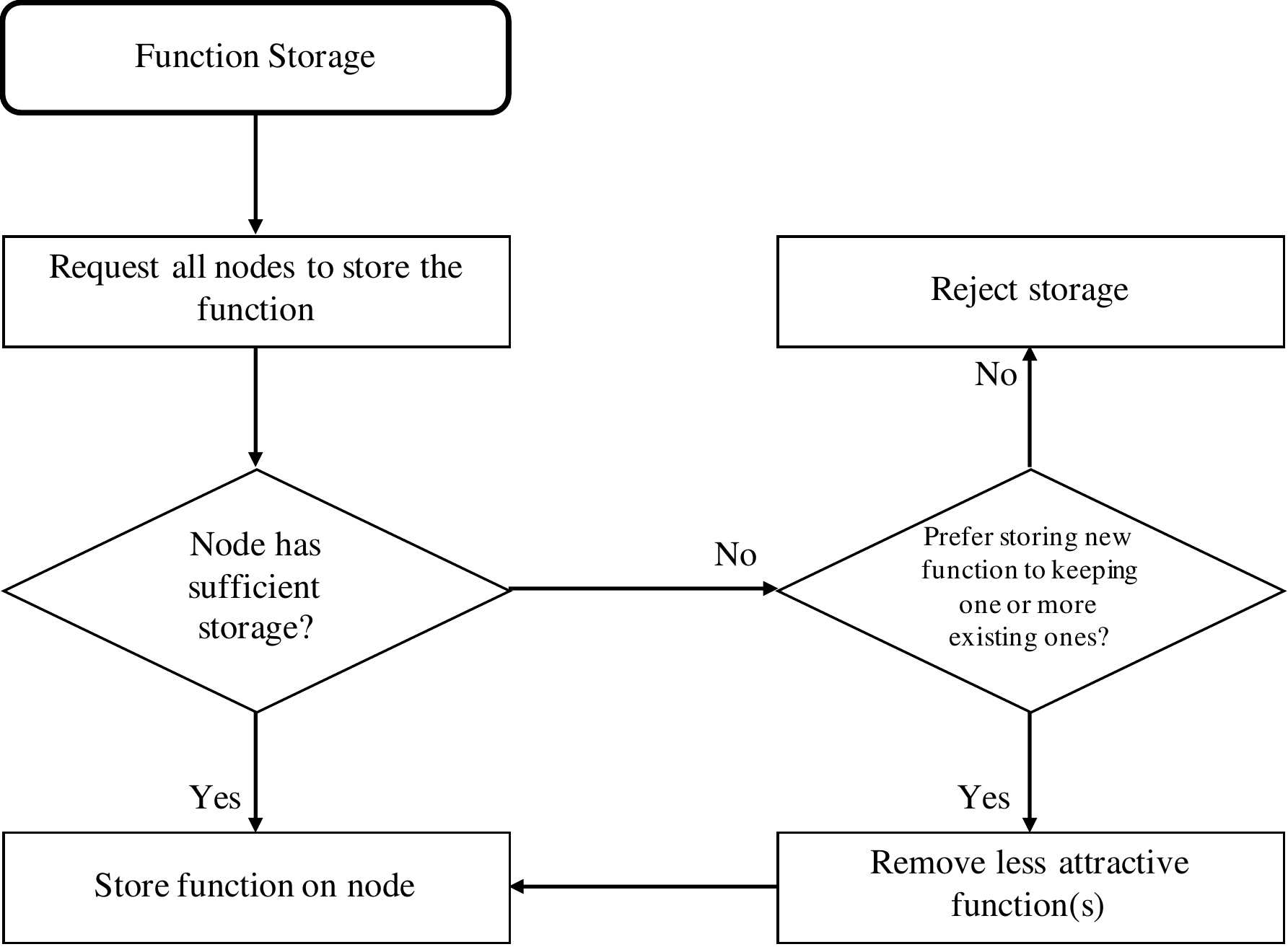}
    \caption{Deployment and storage process of a single function.}
    \label{fig:auction_step1}
\end{figure}

\textbf{Storage Bids:}
The nodes in our approach -- edge, intermediaries, and cloud -- analyze these bids and can make a local decision, i.e., act as auctioneers, which is important for overall scalability and resiliency.
We assume that cloud nodes will accept all bids that exceed some minimum base price.
All other nodes will check whether they can store the executable.
When there is enough remaining capacity, nodes simply store the executable together with the attached bids and start charging the application based on the storage bid.
When there is not enough disk space left, nodes decide whether they want to reject the bid or remove another already stored executable.
For this, nodes try to maximize their earnings.
A simple strategy for this, comparable to standard bin packing, is to order all executables by their storage bid (either the absolute bid or the bid divided by the size of the executable) and remove stored executables until the new one fits in.
Of course, arbitrarily comprehensive strategies can be used (in Section~\ref{subsec:sim3}, we will show through simulation how changing this auctioning strategy affects results) and could even consider the processing bid and the expected number of executions.
In the end, this leads to a situation where some or all nodes will store the executable along with its bids.
See also Fig.~\ref{fig:auction_step1} for an overview of this auction step.

Please note that this process differs from standard auctions since it is not clear upfront how many items are auctioned off, since this depends on the size of the executables as well as the storage capacity of the respective node.

\textbf{Processing Bids:}
When a request arrives, nodes have to decide whether they \emph{want} to process said request.
A node \emph{can} process a request if it stores the corresponding executable and if it has sufficient processing capacity to execute the function.
We can imagine arbitrarily complex schemes for making the decision on whether to execute the request or not if an execution is possible.
For instance, nodes might try to predict future requests (and decide to wait for a more lucrative one) or they might try to queue a request shortly -- all with the goal of maximizing earnings.
In fact, we believe that this opens interesting opportunities for future work.

\begin{figure}
    \centering
    \includegraphics[width=0.6\columnwidth]{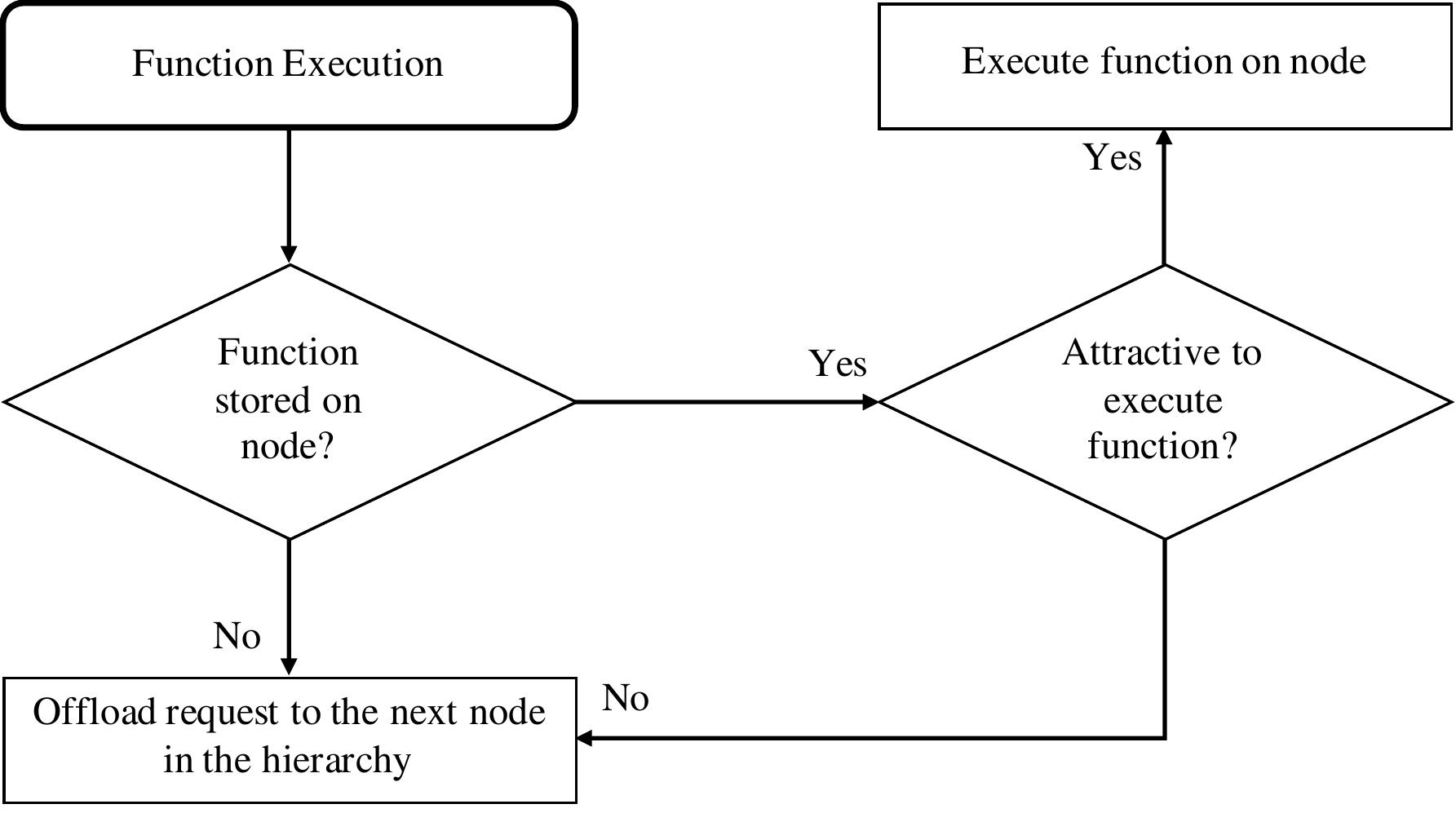}
    \caption{Function execution process for a single node and a single function execution. We assume that the cloud stores and executes all functions with bids beyond a minimum price.}
    \label{fig:auction_step2}
\end{figure}

For the explanations in this paper, we will assume that a node will decide to process a request as soon as it is capable of doing so. This means that upon receipt of a set of incoming requests, a node will follow these four steps:
\begin{enumerate}
    \item Reject all requests for which the executable is not stored locally.
    \item Sort remaining requests by their bid (highest first) into a request list.
    \item While capacity is left, schedule requests from request list.
    \item Reject all requests that exceed the capacity.
\end{enumerate}

Rejected requests are then pushed to the next node on the path to the cloud.
As the cloud -- by definition -- stores all executables and has for all practical purposes unlimited capacity, all requests will at least be served in the cloud.
See also Fig.~\ref{fig:auction_step2} for an overview of this auction step.

Please note that this auction step is different from standard auctions as well in that the number of auctioned items is unknown a-priori.
Instead, it depends on the compute demands and execution duration of requested functions as well as available compute capacity on the respective node.

Overall, this approach solves the question of function allocation, all functions are executed as close as possible to the edge; under high load, it will also have the effect that prices increase towards the edge which is a good incentive considering the high CapEx of installing an edge infrastructure.
Of course, the effects of this allocation scheme also depend on the behaviour and strategies of tenants and nodes in the system, as we discuss in Section~\ref{sec:discussion}.

\section{Challenges for Implementing the Approach in Practice}\label{sec:challenges}
The abstract approach presented in the previous sections has a number of assumptions on FaaS systems which are not necessarily true in existing systems.
In this section, we discuss challenges for implementing the abstract approach in real-world FaaS systems.

We identified four challenges (C1-C4) for practically implementing our idea of an auction-based function placement in real-world FaaS systems. We present each of the challenges first in general, before we discuss the respective challenge using OpenWhisk as an example, as our proof-of-concept prototype (Section~\ref{sec:eval_openWhisk}) is based on OpenWhisk.

\subsection{C1: Auctions Arrive in Batches}
\label{subsec:batch}

In our approach, nodes decide on the execution of specific functions by comparing the execution bids of requests.
For this, we proposed a simple decision process where all incoming requests are ordered from high to low bids and the highest bid gets accepted as the execution set is limited by the required capacity.
This process is sufficient as long as requests arrive in batches periodically.
However, in many real-word scenarios requests will arrive arbitrarily and, therefore, it is not possible to directly compare the execution bids of all incoming requests.
One approach to address this challenge, which we denote as \emph{C1}, is collecting incoming requests for a certain time, effectively creating windows over the incoming stream of execution bids and, thus, the illusion of finite batches.
The downside of this approach is that extra latency (up to the window length) is added to each request in each stage.
Especially, requests that in the end are executed in the cloud will be affected by this extra latency several times.
On the other hand, this approach is very simple to implement.
It is even possible to use windowing only when the local resource utilization exceeds a predefined threshold, i.e., when there is a chance of requests being offloaded.
Another option is the use of statistical methods to estimate the arrival of incoming requests and height of bids, to use the available capacity for requests with relatively high bids based on the prediction model.

\paragraph{OpenWhisk} In its architectural model, OpenWhisk separates load balancing from the part which limits the number of allowed function calls per minute.
Therefore, the decision between execution and offloading has to be shifted to the decision-making part, while the Load Balancer employs strategies to select those functions that optimize the reward.
To implement the two previously described methods, the Load Balancer could simply calculate an average execution bid over a certain timeframe.
Assuming that the node has sufficient free capacities to accept a request, higher bids than the average get accepted while lower bids get offloaded.

\subsection{C2: Storage Space Calculation}
\label{subsec:space}

Our approach assumes that the size of a function's executable is provided along with its storage bid.
It also assumes that executables with lower bids get evicted to free up storage space whenever demand exceeds capacity.
In practice, however, it is difficult to accurately estimate the storage space a function needs.
Nevertheless, handling storage space well is important as the storage on edge devices is often significantly constrained.
Allowing such nodes to store a certain number of functions might be a simple solution, yet the size of functions can differ considerably.
For instance, a simple function may only need a few kilobytes (essentially, the cost for storing a few lines of code in a text file or in memory) while the size of rule-based classifier functions that bundle larger machine learning models may be in the high megabytes.
Consequently, the number of functions is not directly suitable to constrain storage use.
Instead, we need to consider the exact amount of required storage space for functions.
Furthermore, most FaaS systems not only store the functions themselves, but also some data on executions and other metadata, where the size of this information grows over time and also needs to be taken into account.
Arguably, the best approach may be to not only run the auction decision upon deploy time but rather to plan with a safety buffer in terms of storage space and to periodically rerun the auction over the set of locally stored executables and their metadata.

\paragraph{OpenWhisk} OpenWhisk uses a database to store the functions and another one to store the metadata and output of successful executions.
By setting a storage limit, each node can define the storage space available for functions:
Since the exact function size in OpenWhisk is not known before the function is stored in the database, we have to ensure that the free storage space exceeds the maximum function size that we allow OpenWhisk to store by having a sufficiently large storage buffer\footnote{An OpenWhisk version released while we were writing this paper changed this: OpenWhisk no longer checks whether the available storage exceeds the maximum function size. Instead, it checks whether the available storage size exceeds the actual function size plus an estimated buffer based on the base64 encoding in CouchDB. We decided not to change this in the paper text as our prototype fork is based on the previous version and also since it does not really change the overall problem that the precise storage needs are unknown.}.
Moreover, we can use a second buffer to implement a limited storage for the metadata of previously executed functions.
Once this buffer is full, the system can clean the history of execution data to ensure that the node can store data on further executions, e.g., with an LRU eviction scheme.
Thereby, the buffers ensure that enough free space is available to store the functions, past executions, and metadata.

\subsection{C3: Compute Load Estimation}
\label{subsec:load}

Our approach entails that the FaaS nodes can execute a certain number of functions at the same time.
In practice, however, one faces the challenge of having to estimate the supported number of concurrent function invocations.
Setting the number of handled functions $\#\text{func}_{n}$ as shown in Equation~\ref{eq:handled_functions}, where $\text{mem}_{n}$ is the node's available memory and $\text{memreq}_{i}$ the memory requested by function $i \in \{1,2,\dots,m\}$ (as in existing cloud FaaS deployment models), could lead to a simple result.

\begin{equation}
    \label{eq:handled_functions}
    \#\text{func}_{n} = \frac{\text{mem}_{n}}{\frac{1}{m}\times\sum\limits_{i=1}^{m} \text{memreq}_{i}}
\end{equation}

However, especially simple functions often do not fully use their reserved resources.
In addition, most FaaS system support multiple programming languages and Docker containers, so that the resource usage of even similar functions can differ significantly.
Therefore, FaaS systems can overbook the requested memory and often only distinguish between available and unavailable execution machines depending on the components' response times.
In practice, we would suggest to couple the slight overbooking of memory resources with implicit CPU limits based on the number of local cores and the amount of compute capacity allocated to function instances.

\paragraph{OpenWhisk} Instead of using a fixed amount of concurrent functions per Invoker, OpenWhisk distinguishes between healthy and unhealthy Invokers.
However, when deploying OpenWhisk on more restricted hardware with capacities only for a limited number of Invokers, it is important to have a good estimate for the function load.
In addition, an unhealthy Invoker is identified as such after not responding for a certain time, yet this delay can be an issue for low-latency use cases, since latency can grow considerably before the Invoker is switched into an unhealthy state.
Before simply distinguishing between healthy and unhealthy controllers, OpenWhisk used a method that checked the system load through active concurrent invocations in the system.
This would arguably be a better strategy for low-latency use cases but may be problematic when resources are already constrained as on the edge.

\subsection{C4: Knowledge About the Next Node Towards the Cloud}
\label{subsec:next-node}

Our approach assumes that nodes have knowledge about the next nodes towards the cloud.
However, fog and IoT environments are often dynamic and latency between devices changes over time.
In addition, nodes may fail and so can, correspondingly, any specific data transmission route towards the cloud.
Therefore, practical implementations of our approach require a communication mechanism that allows forwarding to the cloud even in situations where the next node toward the cloud is not known and that is also robust against individual node failures.
This could, for instance, be achieved by storing the full path to the cloud on every node rather than the identity of the next node.

\paragraph{OpenWhisk} OpenWhisk was designed for deployment in a single cloud datacenter, not for fog environments.
Therefore, offloading and shifting requests inside a network of OpenWhisk deployments is not part of OpenWhisk yet.
In fact, it may be quite hard to implement since the information necessary for an offloading decision may be distributed over multiple components and machines of the OpenWhisk cluster.
This means that the main challenge for OpenWhisk may be to transfer the abstract concept outlined above into a concrete implementation.

\section{Evaluation: Simulation}\label{sec:eval_sim}
In this subsection, we describe insights we gained from simulation.
Namely, we ran three different simulation experiments:
The first analyzes the effect of processing prices on function placement and latency depending on the request load.
The second studies the effect of storage prices on function placement and latency depending on the demand for storage.
The third experiment comprises a larger simulation in which we study how different executable replacement strategies, that are applied when running out of storage capacity, affect the storage and processing earnings of nodes.

\subsection{Simulation Implementation}

To evaluate our approach, we have implemented a simulation tool in Kotlin which is available on GitHub\footnote{github.com/OpenFogStack/faas4fogsim}.
In the tool, we distinguish three types of nodes: edge, intermediary, and cloud.
All three node types have distinct storage and processing capacities.
Since we assume that all requests require the same compute power but may take arbitrarily long, the processing capacity of a node is specified as the number of requests which can be handled in parallel.
The storage capacity of a node is specified as the number of function executables that the node can store on average.
All nodes treat bids in the way described in Section~\ref{sec:approach}.
For future research on different auction strategies, it is only necessary to change the methods \texttt{offerExecutable} and \texttt{offerRequests} in ComputeNode.kt which implement the respective allocation rules (see Section~\ref{sec:background}).

Beyond the parameters already mentioned, users can specify the node setup (number and type of nodes and their interconnection), average latency for request execution as well as latency between edge and intermediary or intermediary and cloud, average storage and processing bids, the number and average size of executables, as well as the number of requests that should arrive at each edge node.
The simulation comes pre-configured with a set of standard settings (defined in Configuration.kt) that can be changed to customize the simulation for various purposes.
Our implementation also assumes that bids are identical across all node types, i.e., developers will bid and pay the same for execution and storage -- no matter on which node.

To support reproducibility of simulation runs, the tool explicitly sets the random seed; parameters specified as average \textit{X} follow a uniform distribution over a specified range.

\subsection{Configuration}
For the first two simulation experiments, we had one edge, intermediary, and cloud node and simulated a period of two minutes; we also set the average function processing latency as 30ms, the average edge to intermediary latency as 20ms, and the average intermediary to cloud latency as 40ms.
For the third simulation experiment, we extended the topology: it comprised 5 edge nodes per intermediary, 3 intermediary nodes, and one cloud node, i.e., 19 nodes in total.

In the first experiment, we set the storage capacity of each node to a very large value so that function execution was only delegated towards the cloud if a node did not have enough processing capacity available.
All other settings remained as defined in the standard configuration.
As a result, anything up to 166req/s could on average be handled at the edge since the edge could process 5 requests concurrently.
Furthermore, anything up to an average of 832req/s could be handled at edge and intermediary since the intermediary could process 20 requests concurrently.
Anything beyond this required the cloud for handling the load.
Each request had a random processing bid $b_{processing} \sim \mathcal{U}_{[50,150]}$.
We repeated this experiment with varying request load: from 100 to 10,000req/s.
For our 2min (simulated time) experiments, we thus simulated 12,000 to 1,200,000 requests.

In the second experiment, we set the processing capacity of each node to a very large value so that function execution was only delegated towards the cloud if a node did not have enough storage capacity to store the function executable.
All other settings remained as defined in the standard configuration.
As a result, edge nodes could on average store 10 functions, intermediaries 50 functions, and the cloud an unlimited amount of functions.
Executable sizes were randomly chosen as $size \sim \mathcal{U}_{[0.5,1.5]}$.
Each executable also had a random storage bid  $b_{storage} \sim \mathcal{U}_{[50,150]}$.
We repeated this experiment with varying numbers of executables starting with 5 (which should still all fit on the edge node), over 50 (on average, edge nodes stored 20\% of all executables and intermediaries still stored most of them), up to 100 in steps of 5.

In our third experiment, we evaluated the impact of our simple node strategy on function execution.
As we have discussed, non-cloud nodes make their decision about which function executables to accept or drop based solely on storage bids and do not consider the processing bid.
We added a \emph{stickiness} parameter to the simulation that describes the probability of a node ignoring an incoming function executable with a higher storage bid in favor of keeping its current set of function executables:
A stickiness value of 0 corresponds to the default auction case from the other two simulation experiments, a stickiness value of 0.5 means that a bid which would normally evict other executables will do so with only a 50\% probability, a stickiness value of 1 means that a node will never evict already stored executables (it will ``stick'' with the already stored ones).
Please, note that we do not consider such a stickiness-based approach as a good strategy (as in ``randomize if you follow the auction rules or not'').
Instead, different strategies, e.g., based on local preferences regarding certain users, will lead to different outcomes in terms of stickiness -- hence, this is only a parameter to capture implications of a broad range of strategies as part of our simulation model, i.e., variability of the sets of stored executables.
For this experiment, we set the number of executables to 100.
All other settings remained as defined in the standard configuration except for our updated topology (19 instead of 3 nodes).
As a result, executable sizes were randomly chosen from $\mathcal{U}_{[0.5,1.5]}$, which means that edge nodes stored on average 10\% and intermediaries stored on average 50\% of all executables.

\subsection{Simulation Experiment 1: Effect of Processing Prices}

\begin{figure}
    \centering
    \begin{subfigure}{0.6\columnwidth}
        \centering
        \includegraphics[width=\columnwidth]{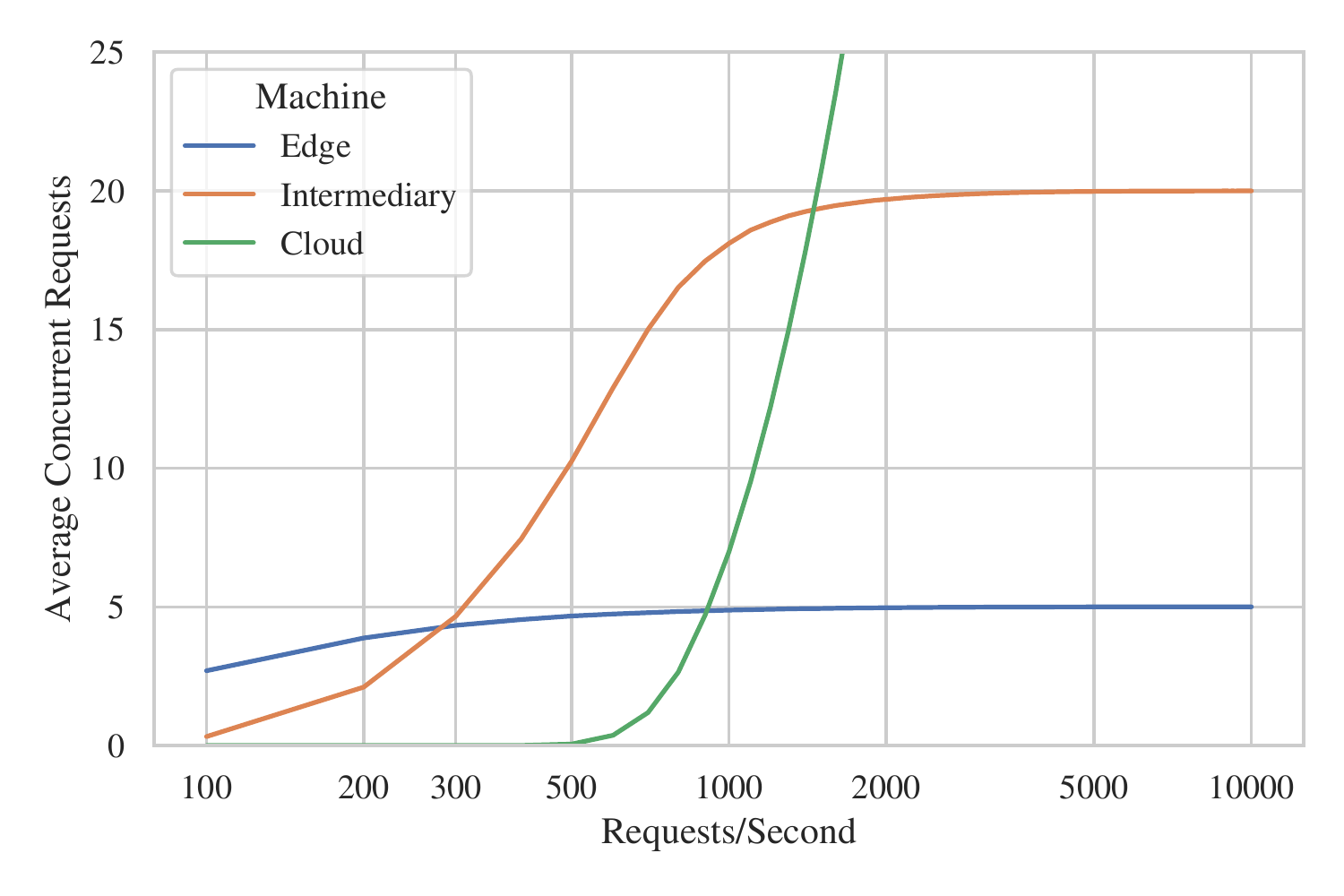}
        \caption{Number of concurrent function executions}
        \label{fig:exp1-concurrent_requests}
    \end{subfigure}
    \hfill
    \begin{subfigure}{0.6\columnwidth}
        \centering
        \includegraphics[width=\columnwidth]{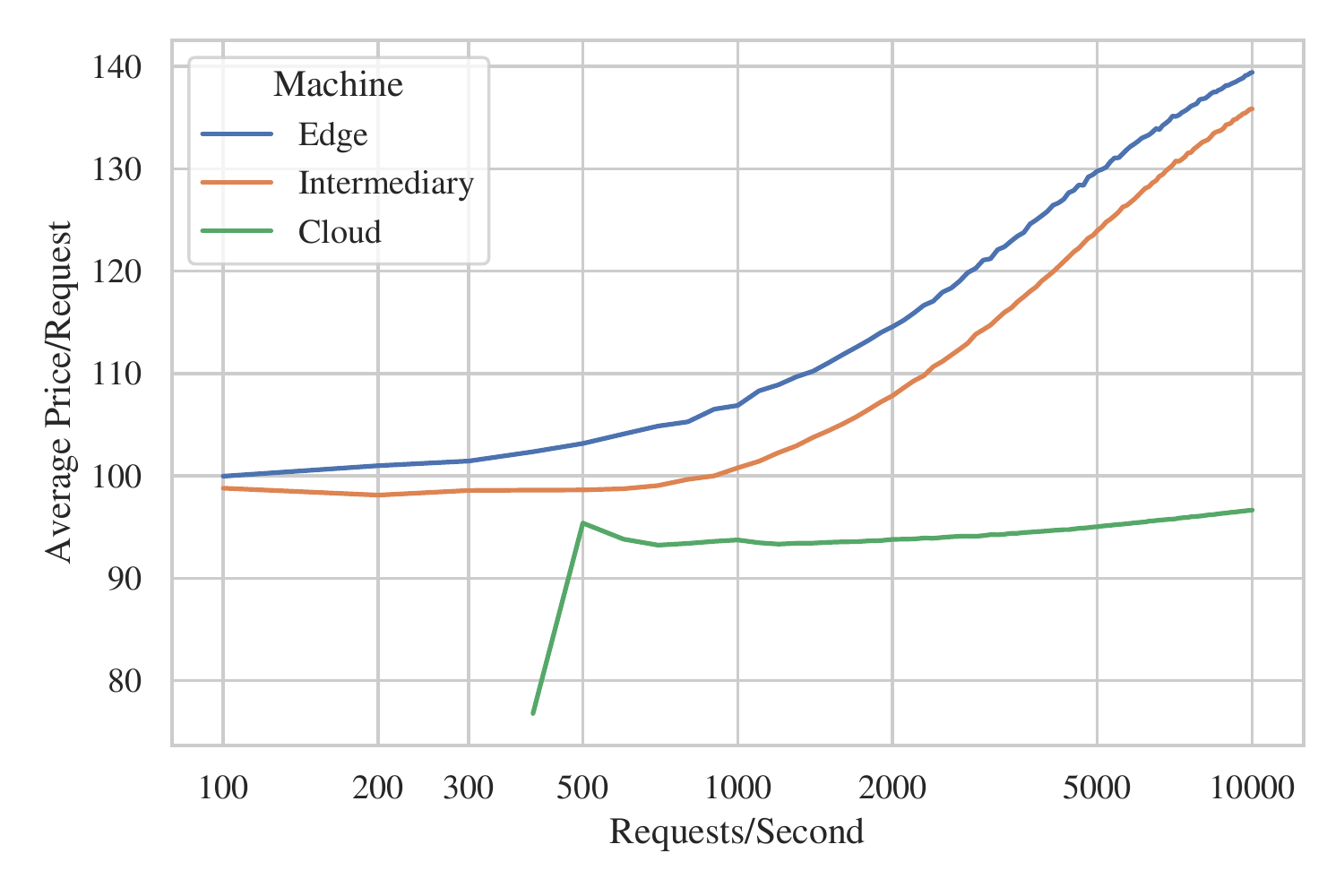}
        \caption{Average price per function execution}
        \label{fig:exp1-price}
    \end{subfigure}
    \hfill
    \begin{subfigure}{0.6\columnwidth}
        \centering
        \includegraphics[width=\columnwidth]{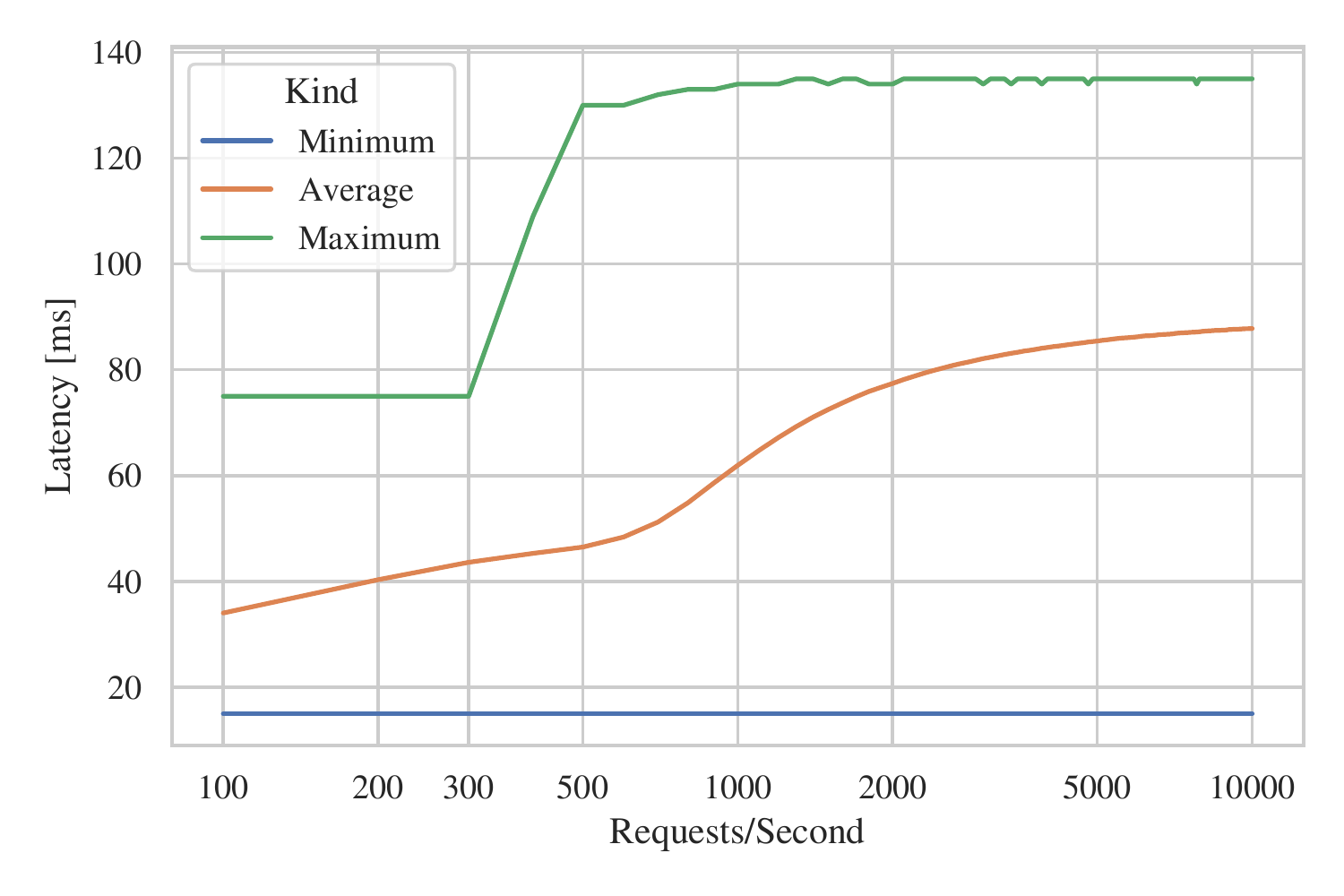}
        \caption{Request latency}
        \label{fig:exp1-latency}
    \end{subfigure}
    \caption{Simulation Experiment 1: Studying the effect of varying request load when only execution capacities are auctioned.}
    \label{fig:exp1}
\end{figure}

As can be seen in Figure~\ref{fig:exp1-concurrent_requests}, the simulated system is lightly loaded up to about 400req/s, since no requests are delegated towards the cloud.
The intermediary begins to handle more requests than the edge at 300req/s and the cloud begins to handle more requests than the intermediary at 1500req/s.

For the earnings, we can see in Figure~\ref{fig:exp1-price} that the average price for executing a function at the edge gradually increases as the edge node selects only the most lucrative requests for execution if it has a choice.
For the intermediary, the average price remains slightly below 100 until 900req/s, then it also starts to gradually increase.
The cloud only processes its first 11 requests at 400req/s.
Since the intermediary only forwards the requests with the lowest processing bids, the cloud earnings are also quite low for this request rate.

For latency, we can see in Figure~\ref{fig:exp1-latency} that it linearly increases with the load as more and more requests (the lower paying ones) are no longer served on the edge but rather delegated to intermediary and cloud.
We can also see the bump in maximum latency when the first request is executed in the cloud.

Overall, this experiment shows that even the simple auction scheme that we implemented is indeed an efficient mechanism to handle function allocation across fog nodes with the goal of determining the placement based on payment preferences of application developers.

\subsection{Simulation Experiment 2: Effect of Storage Prices}

\begin{figure}
    \centering
    \begin{subfigure}{0.6\columnwidth}
        \centering
        \includegraphics[width=\columnwidth]{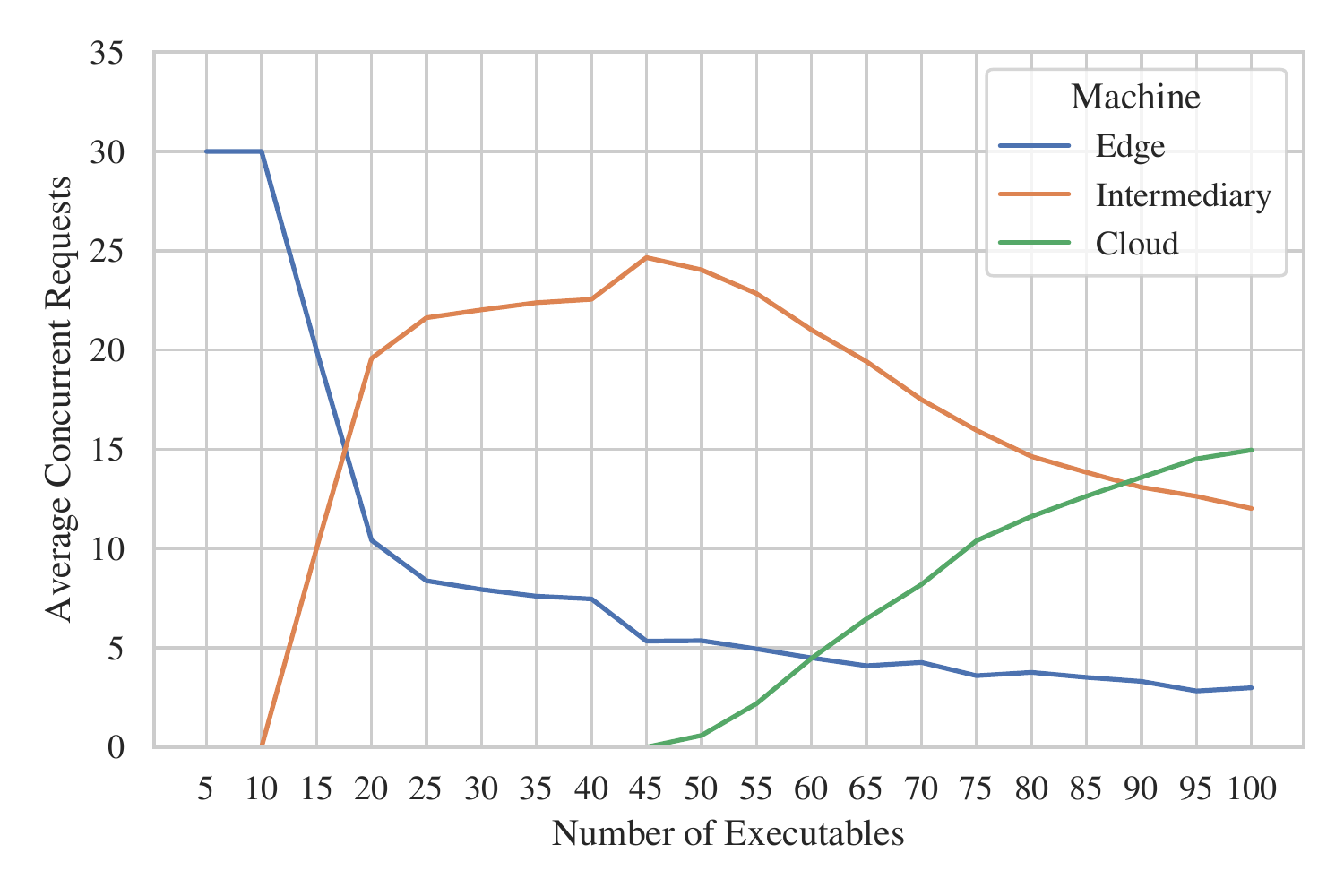}
        \caption{Number of concurrent function executions}
        \label{fig:exp2-concurrent_requests}
    \end{subfigure}
    \hfill
    \begin{subfigure}{0.6\columnwidth}
        \centering
        \includegraphics[width=\columnwidth]{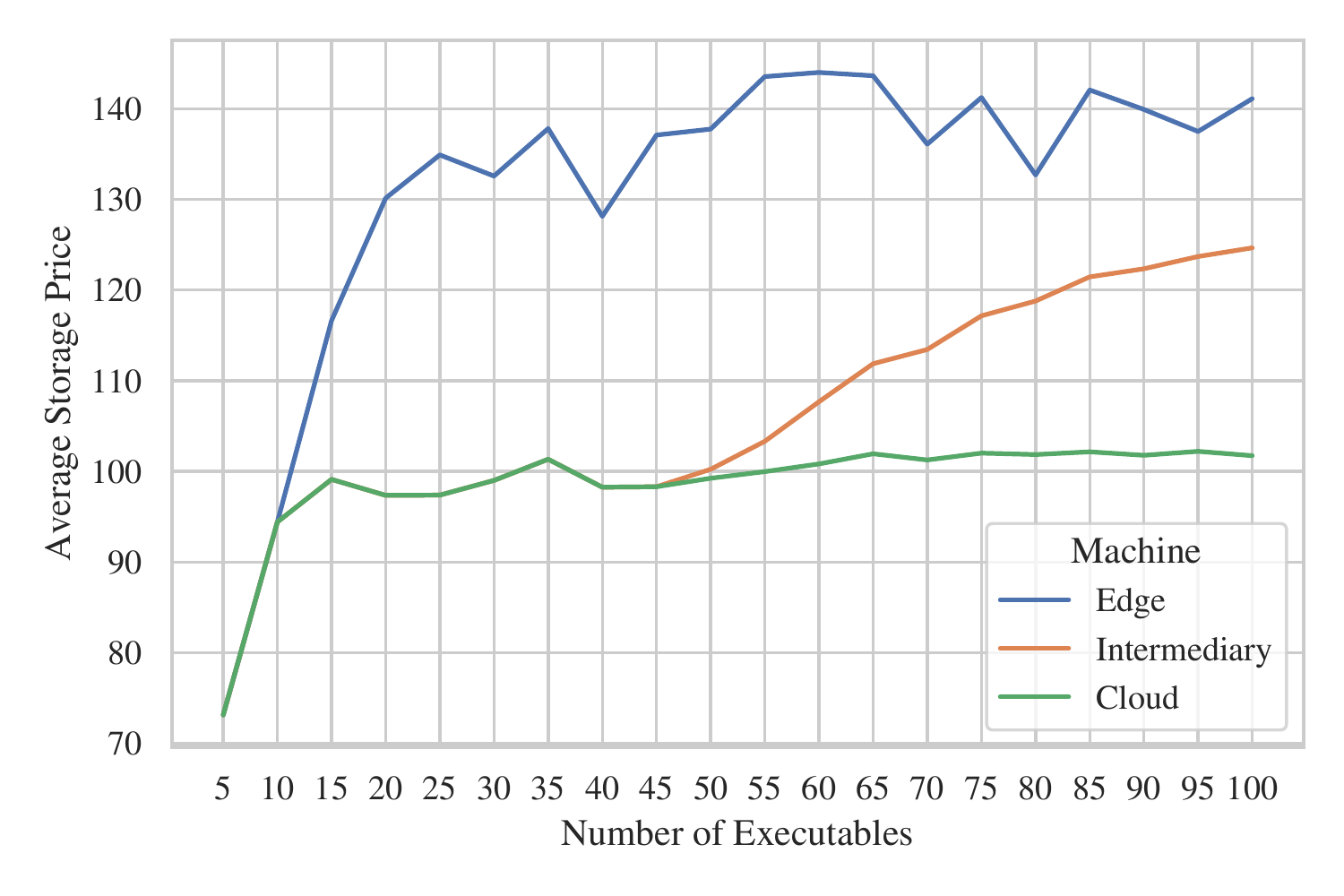}
        \caption{Average storage price per stored function executable}
        \label{fig:exp2-price}
    \end{subfigure}
    \hfill
    \begin{subfigure}{0.6\columnwidth}
        \centering
        \includegraphics[width=\columnwidth]{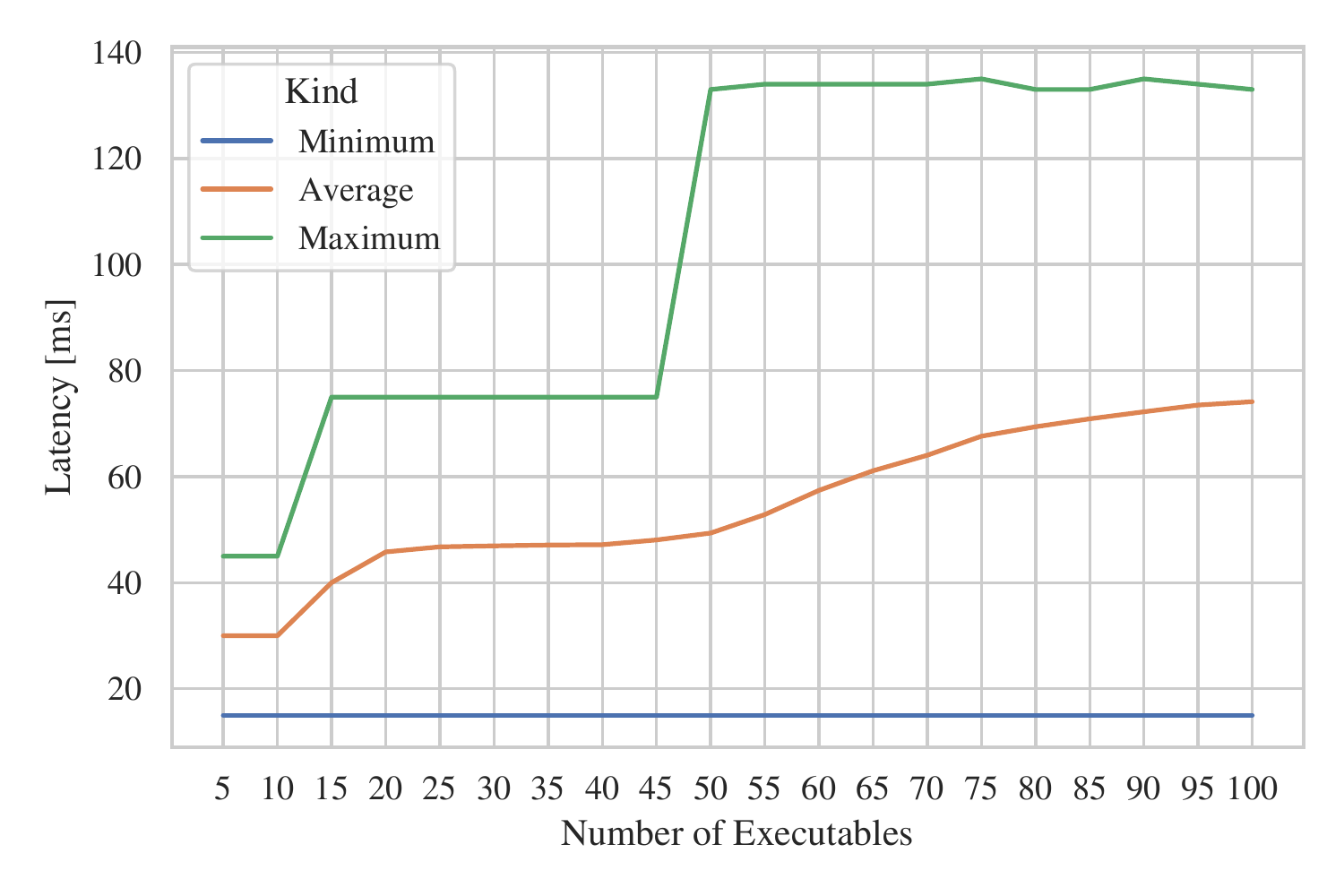}
        \caption{Request latency}
        \label{fig:exp2-latency}
    \end{subfigure}
    \caption{Simulation Experiment 2: Studying the effect of varying number of executables when only the storage capacity is auctioned.}
    \label{fig:exp2}
\end{figure}

In this experiment, we analyze what happens when the number of distinct functions increases (far) beyond the number of executables that can be stored on the edge and later even the intermediary.
As can be seen in Figure~\ref{fig:exp2-concurrent_requests}, the edge node begins to forwards requests to the intermediary at about 10 executables, while the intermediary begins to forwards requests to the cloud at about 50 executables (which is as expected based on our simulation parameters).
The more functions, for which no executable is stored locally, are requested, the more requests are offloaded towards the cloud.

This is also visible from the average storage price shown in Figure~\ref{fig:exp2-price}: the edge does not have any choice of executables when only five are offered so it also has to accept low storage bids.
Likewise, the intermediary only starts to evict executables once it exceeds its capacity of about 50.
Afterwards, both curves gradually increase towards the upper limit of 150 which, however, is not reached in this experiment as the randomization means that only 1\% of the executables will be offered at a storage bid of 150.
In addition, the average storage price in the cloud is about 100 which is as expected as the cloud accepts all executables in our simulation setting.

Figure~\ref{fig:exp2-latency} shows how an increasing number of executables also leads to increasing request latency:
With every additional executable, the share of matching executables available at the edge (or on the intermediary) decreases so that requests are increasingly delegated towards the cloud since the function can, without an executable, not be executed.

Overall, this experiment shows that our auction-based approach is an efficient mechanism to determine distribution of binaries across a fog network when there is a lack of storage capacity.
Based on the experiment results, we would also argue that both bids should not be considered separately: From the perspective of an edge node, storing only the \textit{n} executables that have the highest storage bid may not be optimal. For instance, if the \textit{(n+1)}st executable (in terms of storage bid) is invoked significantly more often than the \textit{n}th, storing the \textit{(n+1)}st instead of the \textit{n}th may offer higher earnings.
The optimum, here, depends on request rates, execution bids, storage bids, and their respective distributions.

\subsection{Simulation Experiment 3: Larger Deployment and Analysis of the Effect of Stickiness in Executable Management\label{subsec:sim3}}

\begin{figure}
    \centering
    \begin{subfigure}{0.6\columnwidth}
        \centering
        \includegraphics[width=\columnwidth]{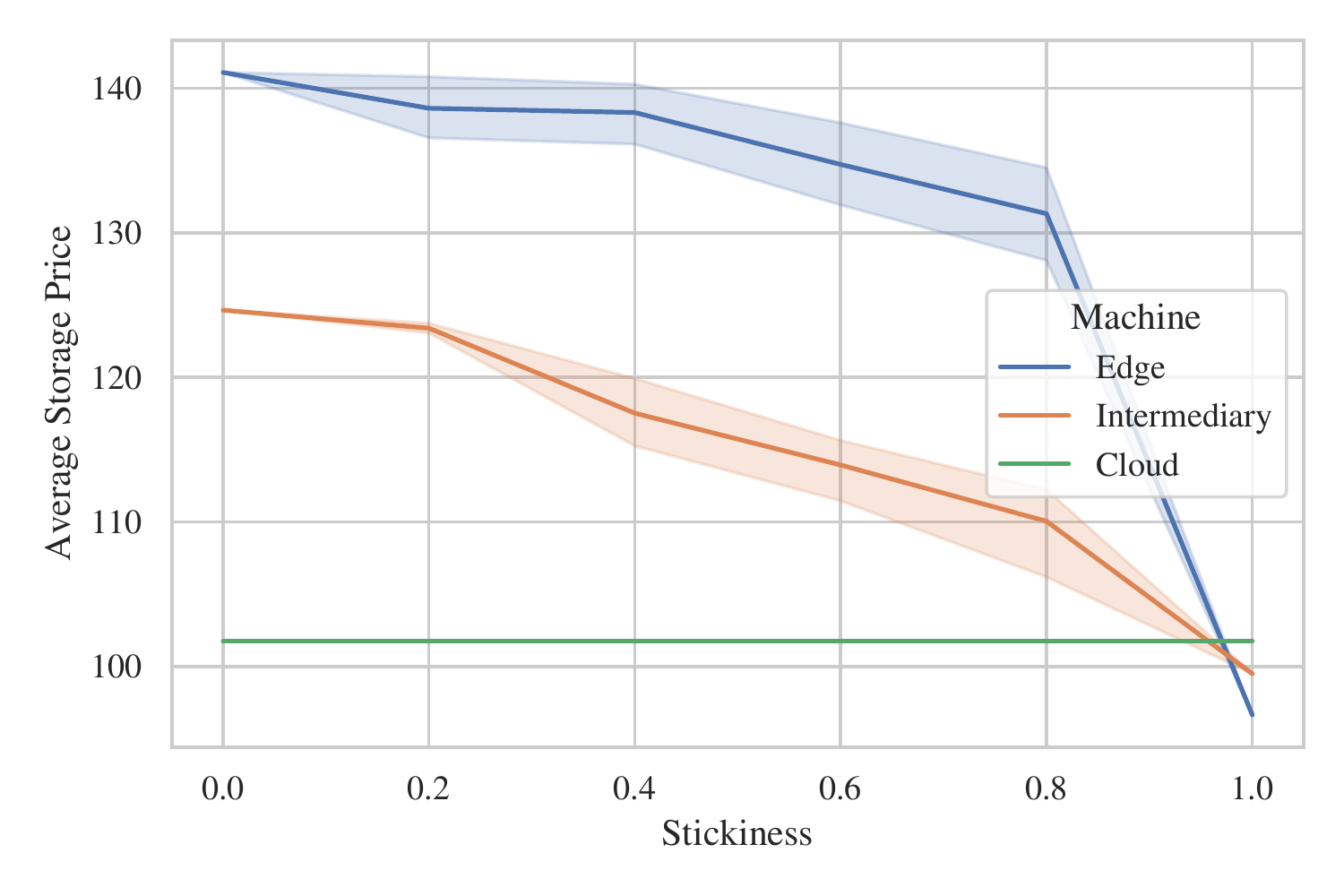}
        \caption{Average storage price per stored function executable}
        \label{fig:exp3-avg_storage_price}
    \end{subfigure}
    \hfill
    \begin{subfigure}{0.6\columnwidth}
        \centering
        \includegraphics[width=\columnwidth]{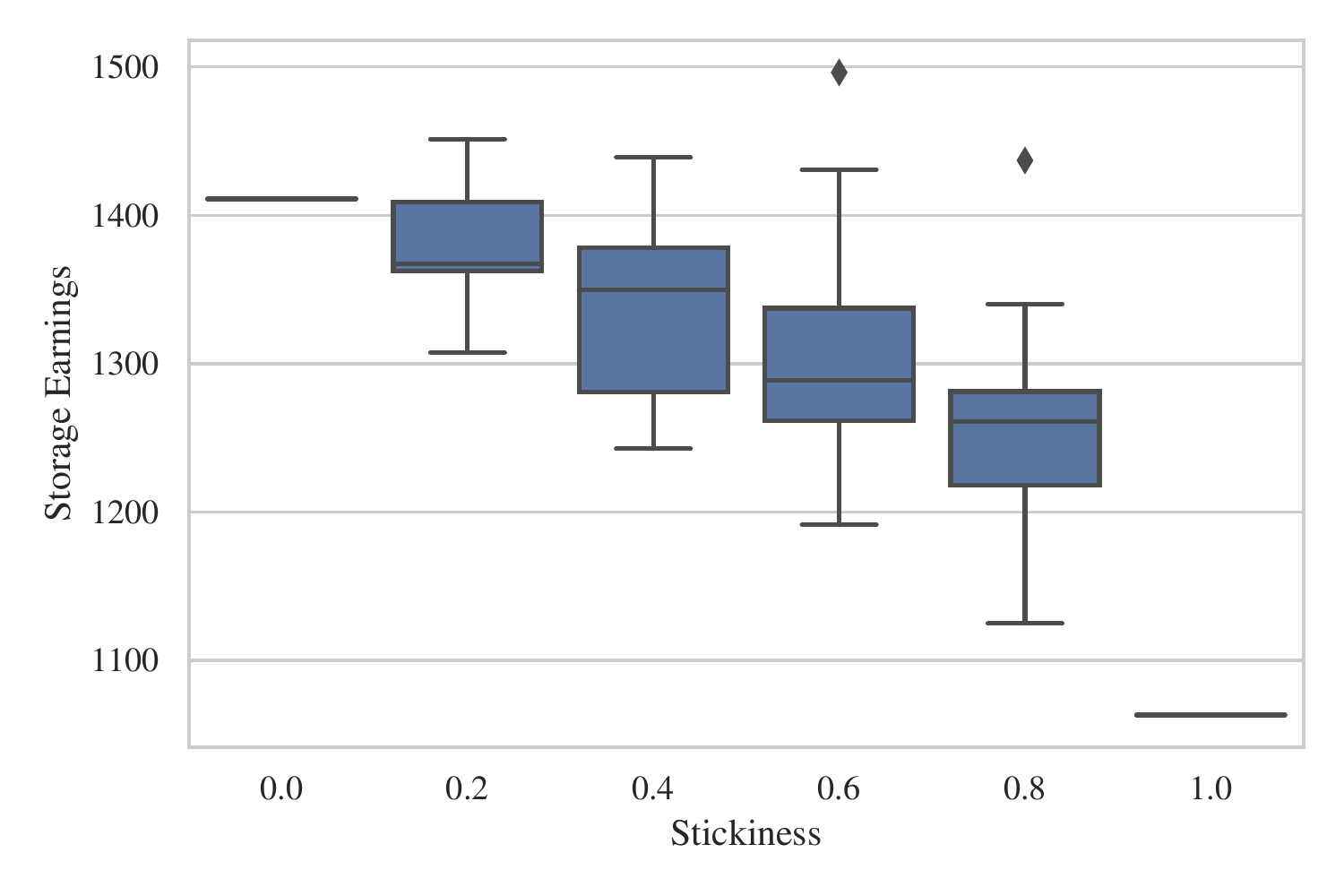}
        \caption{Distribution of storage earnings across Edge nodes}
        \label{fig:exp3-storage_earnings_edge}
    \end{subfigure}
    \hfill
    \begin{subfigure}{0.6\columnwidth}
        \centering
        \includegraphics[width=\columnwidth]{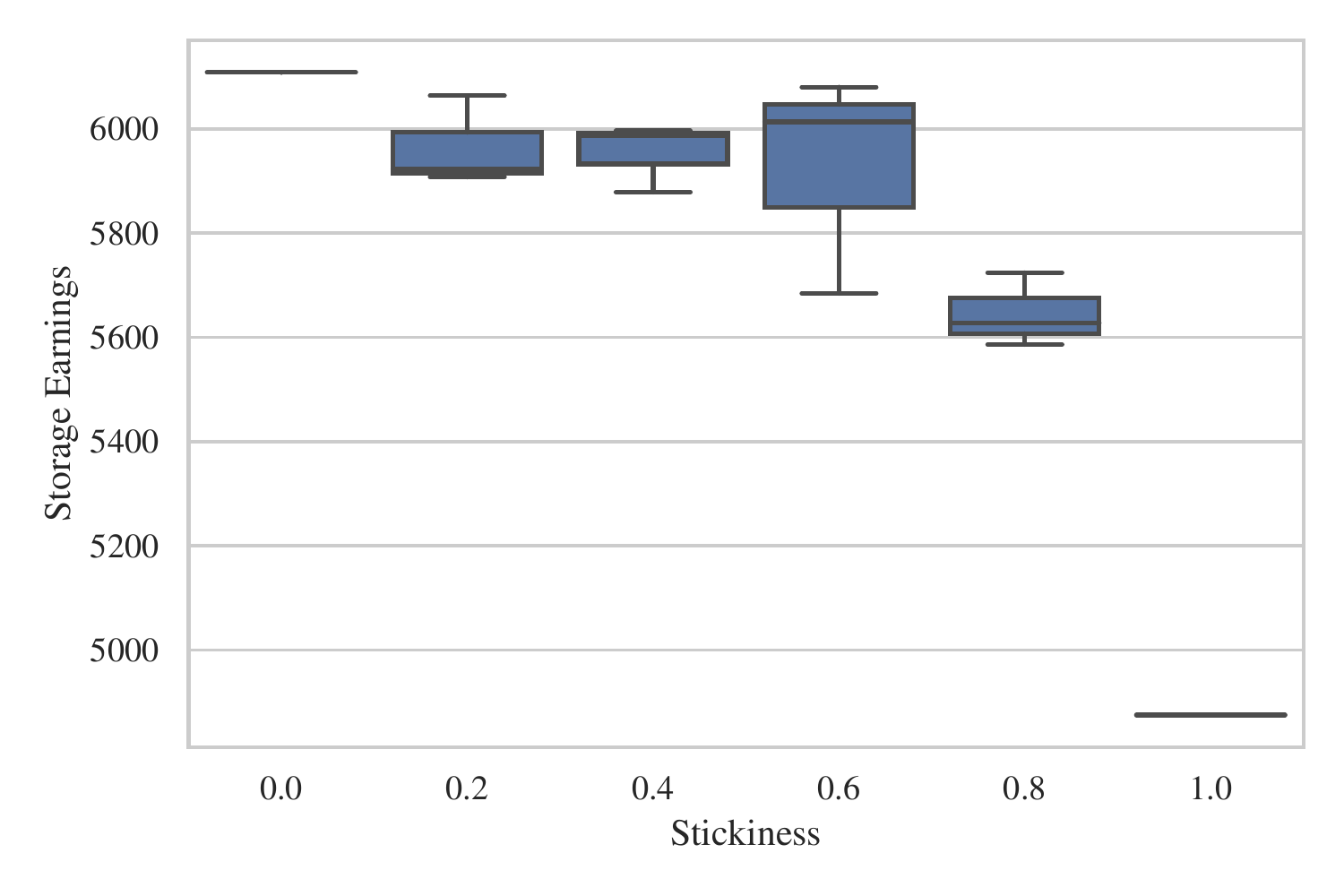}
        \caption{Distribution of storage earnings across intermediary nodes}
        \label{fig:exp3-storage_earnings_intermediary}
    \end{subfigure}
    \caption{Simulation Experiment 3: Studying the effect of varying degrees of stickiness in executable storage on storage earnings (one simulation run).}
    \label{fig:exp3-storage}
\end{figure}

In this experiment, we show that our simulation model is also suited for evaluating larger node topologies and analyze how different node strategies affect storage and processing earnings.
As can be seen in Figure~\ref{fig:exp3-avg_storage_price}, the average storage price for edge and intermediary nodes increases with a lower function stickiness.
The reason for this is that a stickiness level of 0 means that the node will store the top X highest paying executable (in our example: on average the top 10 and top 50 out of 100 for edge and intermediary nodes respectively).
A stickiness level of 1 means that the node will store the \textit{first} X executables that are offered -- essentially a random subset depending on the order in which storage requests come in\footnote{In our simulations setup, storage requests were sent in ascending order ordered by bid. Hence, the stickiness level of 1 results in the minimum possible storage earnings for the respective node.}.
In general, it is therefore best to strictly follow our auction approach and to replace low-bid executables for maximizing the earnings from executable storage.
There are, however, situations in which a node might benefit from not replacing an executable.
For example, not replacing two small executables with a single large executable with a higher bid can result in higher total storage earnings.
This can also be seen in Figure~\ref{fig:exp3-storage_earnings_edge} which shows that some edge nodes achieved higher storage earnings when ignoring 20-80\% of storage requests compared to accepting all requests.
Due to the larger overall storage capacity, this phenomenon does not matter for intermediary nodes in our simulation setup (Figure~\ref{fig:exp3-storage_earnings_intermediary}).

\begin{figure}
    \centering
    \begin{subfigure}{0.6\textwidth}
        \centering
        \includegraphics[width=\columnwidth]{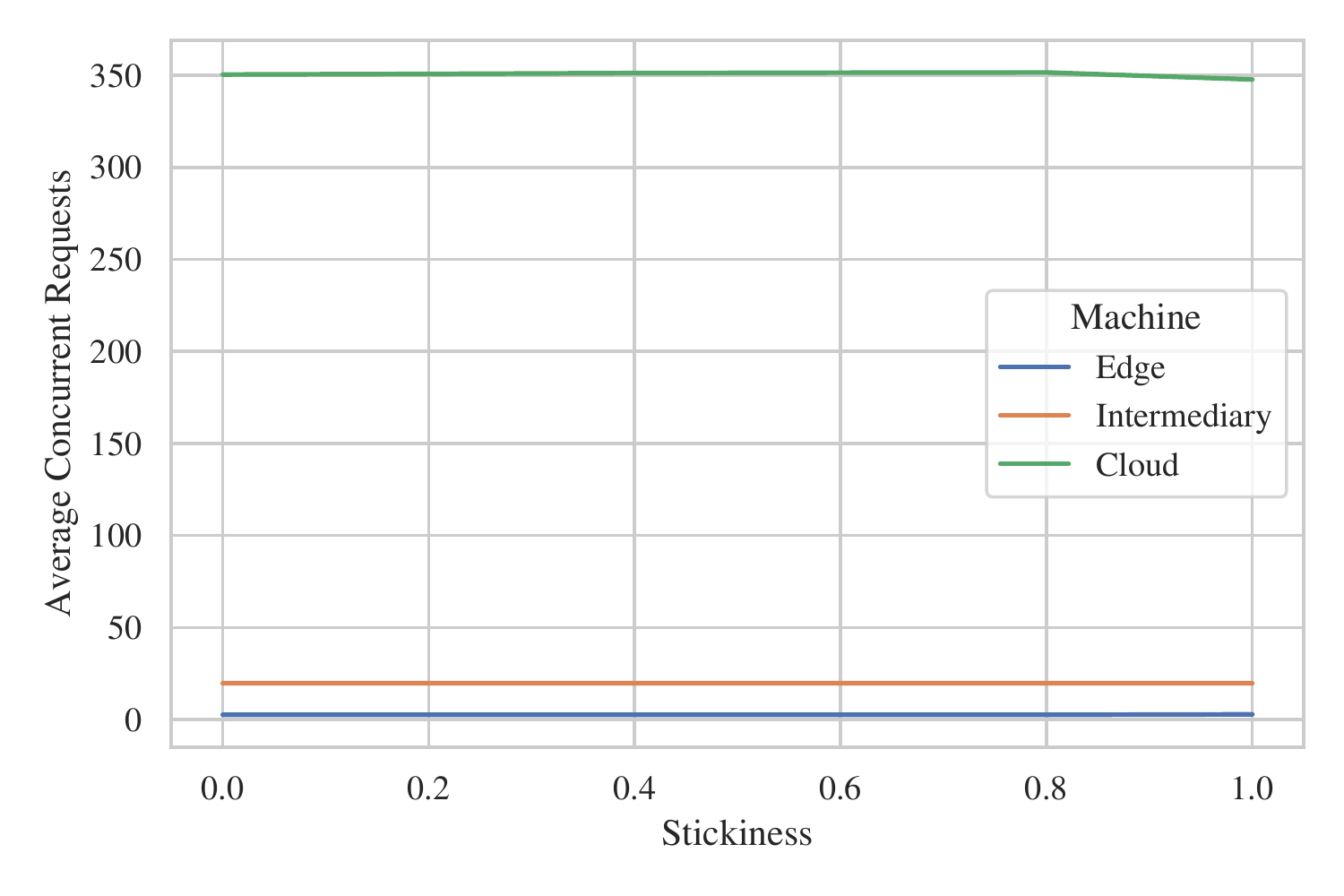}
        \caption{Number of concurrent function executions}
        \label{fig:exp3-concurrent_requests}
    \end{subfigure}
    \hfill
    \begin{subfigure}{0.6\textwidth}
        \centering
        \includegraphics[width=\columnwidth]{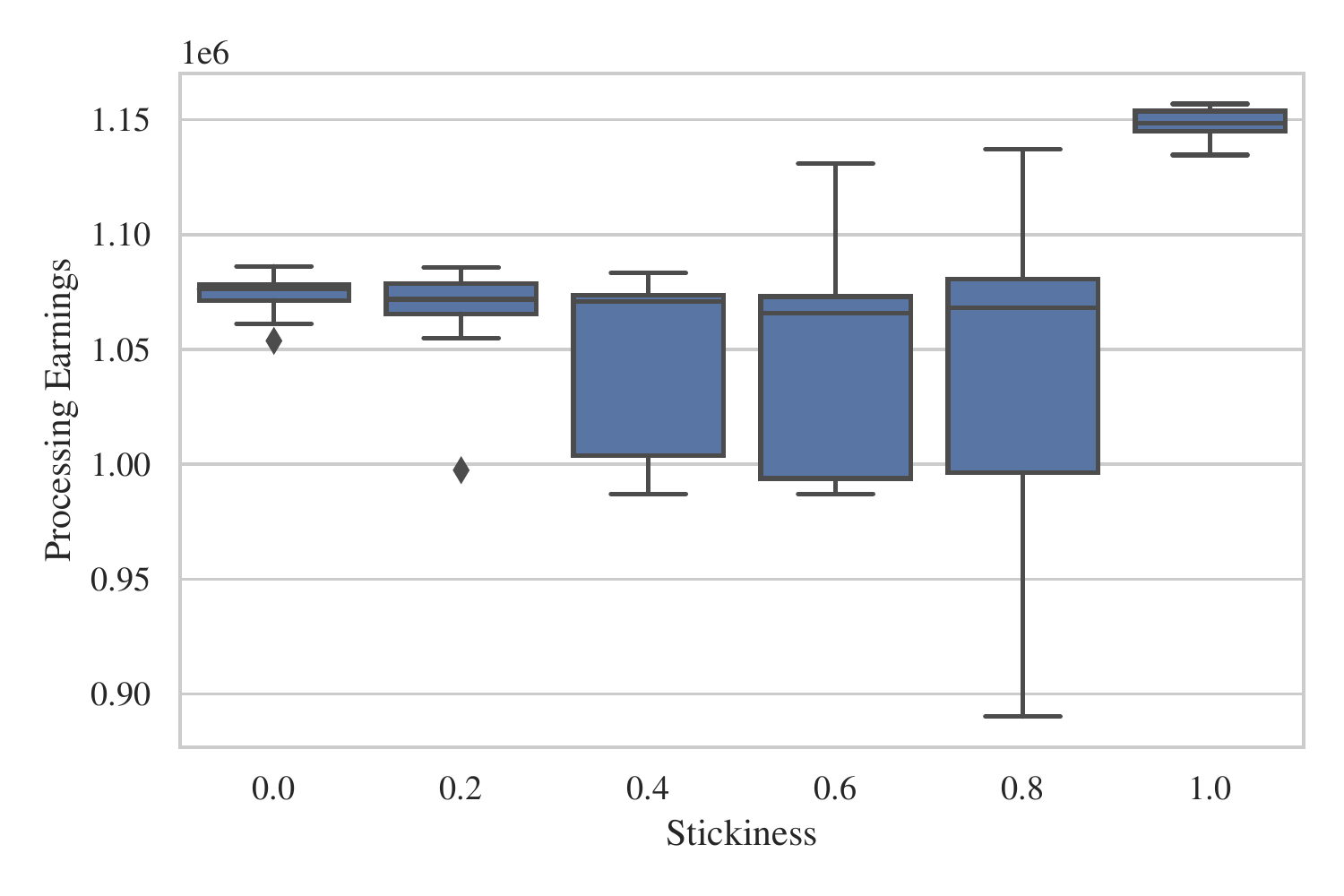}
        \caption{Processing earnings for edge nodes}
        \label{fig:exp3-processing_earnings_edge}
    \end{subfigure}
    \hfill
    \begin{subfigure}{0.6\textwidth}
        \centering
        \includegraphics[width=\columnwidth]{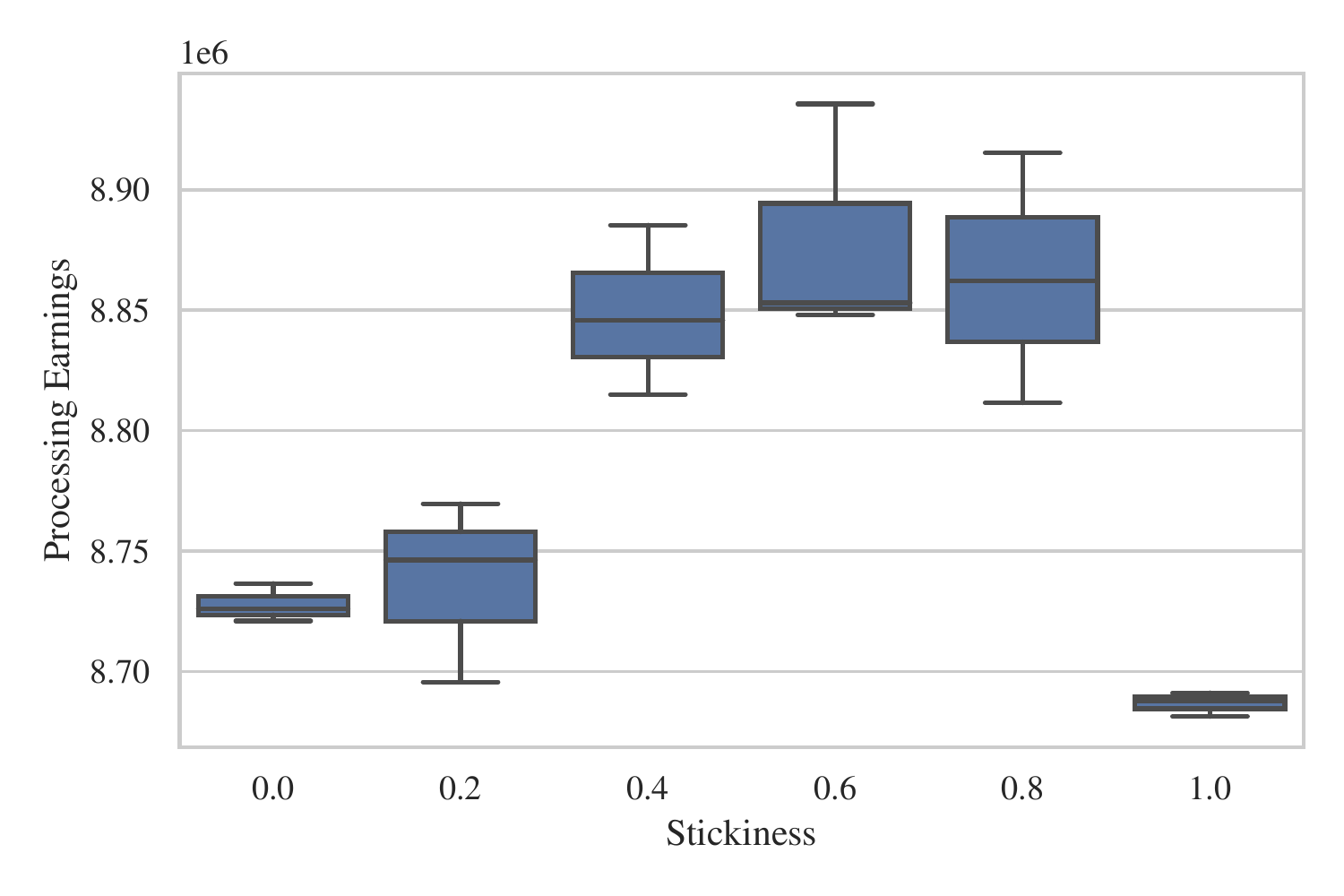}
        \caption{Processing earnings for intermediary nodes}
        \label{fig:exp3-processing_earnings_intermediary}
    \end{subfigure}
    \caption{Simulation Experiment 3: Studying the effect of varying function stickiness on processing earnings (one simulation run).}
    \label{fig:exp3-processing}
\end{figure}

Figure~\ref{fig:exp3-concurrent_requests} shows that function stickiness has virtually no effect on the average number of concurrent requests carried out by each node; edge nodes have a load of 50\% on average while intermediary nodes are fully loaded.
The reason for this effect is that requests follow a uniform distribution across all executables.
For stickiness levels of 0 and 1, all edge nodes (and intermediary nodes likewise) store the same set of executables.
For values in between 0 and 1, there is some variance across nodes as some edge nodes (and intermediary nodes likewise) execute a few more requests.
Due to the uniform distribution of requests, this variance is, however, so low that the curves which show it (using the same formatting as Figure~\ref{fig:exp3-avg_storage_price}) are not visible in the chart.

Furthermore, Figure~\ref{fig:exp3-processing_earnings_edge} shows that the median processing earnings are very similar for all nodes that periodically ignore storage requests since there is no correlation between processing and storage bids in our simulation setup.
A higher function stickiness can lead to higher earnings for edge nodes, albeit at increased risk, as it depends on the processing bids for the two executables which are supposed to be swapped (or not).
The biggest gamble for an edge node is a stickiness of 1: we repeated the experiment with different seeds and the processing earnings for such edge nodes can be very high or low compared to all other outcomes.
The reason for this is that a stickiness level of 1 essentially picks a random subset of executables which each have random processing bids, thus, resulting in fully random processing earnings.

The low variability levels for both stickiness levels of 0 and 1 in Figures~\ref{fig:exp3-processing_earnings_edge} and~\ref{fig:exp3-processing_earnings_intermediary}, where all nodes of the respective type store the same executable set, result from the variability of requests across different nodes.
It can be expected to asymptotically converge to 0 for very long running simulation experiments due to the uniform distribution of requests across executables and nodes.

Figure~\ref{fig:exp3-processing_earnings_intermediary} shows that for intermediary nodes it is a beneficial strategy to choose a stickiness parameter between 0 and 1.
The reason for this is that, in our setup, intermediary nodes maximize the probability of storing a different set of executables than their corresponding edge nodes.
This means that requests offloaded by the edge due to a missing executable are more likely to find that executable on the intermediary (in the best case in our setup around 50\% rather than 40\% probability), i.e., the executable has a choice from more requests and can achieve higher processing earnings.
This effect is less pronounced when edge nodes are overloaded.

Overall, this shows that nodes should consider both processing and storage bids in an optimal strategy.
They should also consider the frequency of execution requests for different executables.
Finally, intermediary nodes should closely monitor the strategies of their edge nodes (or other intermediary nodes which are child nodes of them) and adapt correspondingly.
For this, it may suffice to pretend that their respective child nodes are end devices sending requests and, otherwise, follow the same workload prediction strategies as edge nodes.

\section{Evaluation: Prototype and Experiments}\label{sec:eval_openWhisk}
In this section, we present our proof-of-concept prototype AuctionWhisk which implements our auction-based approach in the open source FaaS platform (Lean) OpenWhisk.
We then evaluate our approach through a number of experiments with AuctionWhisk.

\subsection{Prototypical Implementation of AuctionWhisk\label{sec:prototype}}

Our AuctionWhisk prototype is based on OpenWhisk and also supports the Lean OpenWhisk deployment type, which we refer to as Lean AuctionWhisk.
Our auction-based function placement approach consists of two rounds, we therefore decided to split the description of our implementation into two parts:
First (Section~\ref{subsec:store}), we describe how to handle the storage bids and the resulting auction round.
Second (Section~\ref{subsec:execution}), we present how we manage the execution of functions based on auction bids.
We have made AuctionWhisk available as open source\footnote{github.com/OpenFogStack/AuctionWhisk}.

\subsubsection{Storing Function Executables}
\label{subsec:store}
Before a function can be called, the executing node has to store the executable locally.
OpenWhisk relies on CouchDB for storing function executables\footnote{OpenWhisk refers to functions as \emph{actions}.}.
In this context, there are two relevant CouchDB databases: \texttt{whisk\_local\_whisks} stores the function executable and \texttt{whisk\_local\_activations} stores information about successful executions, including output and metadata.
In AuctionWhisk, we extended the data model of \texttt{whisk\_local\_whisks} to also store the bids along with the executables and adapted the components interacting with this data model to also pass on the additional data.

In Section~\ref{subsec:space}, we discussed challenge C2 and possible solutions for how nodes can estimate the required storage.
In AuctionWhisk, each node can define the storage space available for functions by setting a configuration parameters for the per-function storage limit as well as the total storage space available to AuctionWhisk.
Both limits include all metadata of executed functions in \texttt{whisk\_local\_activations} which therefore should be garbage-collected periodically.

As a limitation of the original OpenWhisk, the actual storage size of a function is unknown.
This also means that the Controller, which decides on accepting or rejecting a function executable, cannot directly know the actual amount of used storage.
To circumvent this problem, we decided to use the per-function storage limit (which by default in OpenWhisk is 48MB) as a proxy for the actual function size.
Therefore, when AuctionWhisk is offered a new function executable, the system calculates the currently used storage as the number of stored executables times the per-function storage limit and compares it to the total storage limit.
If the estimated free space is larger than the per-function storage limit, AuctionWhisk accepts the new executable and inserts it into the database.
If not, AuctionWhisk retrieves the executable with the lowest storage bid and compares the bid to the new executable's bid.
The one with the higher bid is then stored.

While this means that we reserve the per-function storage limit for every function, even for those that are possibly only a few KB in size, we believe that this is an elegant solution to the problem:
The decision of which executables to store can easily be made and storage needs to be planned with a safety buffer anyhow since any execution of the function will increase the amount of stored data and functions may also need temporary disk space while executing.
For a production environment, one could also analyze the actual function size after the deployment and use that for calculating the currently used storage for later requests. We decided not to do this in our prototype as this would require significant modifications to the OpenWhisk code base.

\subsubsection{Executing Function Calls}
\label{subsec:execution}

Upon invocation, OpenWhisk has to decide whether to execute or to offload the request based on the execution bid.
This challenge is discussed in Section~\ref{subsec:next-node} as C4, where we assume that the nodes have knowledge about the next node towards the cloud.
Once a request arrives at our node, the system checks whether the function executable is available locally.
If this is not the case, standard OpenWhisk would return an error.
In AuctionWhisk, however, the request is in that case forwarded to the next node on the path towards the cloud.
The downside of this approach is that requests that target a function that does not exist anywhere in the system will always get their (high-latency) error message from the cloud.
We believe, however, that this is acceptable since (i) we expect this scenario to be rare and (ii) the benefit of not having to store information on all functions on every FaaS node far outweigh this disadvantage.
For instance, FaaS nodes in the US do not have to be aware of functions relevant only in Australia.
We implemented this offloading behavior as part of the Controller and forward the request directly to the next node -- ideally, an external messaging system would take care of delivering the request to the next \emph{available} node on the path to the cloud.

As discussed with C3 (Section~\ref{subsec:load}), it is relatively hard to estimate the available compute capacity, i.e., how many additional requests can be accepted.
Since OpenWhisk already uses the number of concurrent function executions to estimate CPU load, we have decided to adapt this approach in AuctionWhisk.
We introduce a configurable parameter that allows users to define a time window length as well as the number of parallel requests that can be accepted in that time window.
This is similar to a previous implementation in OpenWhisk but can be decided in the load balancer part of the Controller as it does not require information from all invokers of the node.
If the request rate and the kind of function requested stays relatively constant, this approach allows a platform operator to find a good combination of configuration parameters.

If the average duration of requested functions increases, requests will be queued briefly; if it decreases requests will be offloaded even though there is still capacity left.
On the one hand, this solution is simple to implement as it only requires some modifications to the existing code base.
It also allows for a more fair and fine-grained scheduling of resources by the operating system scheduler as processes can requests all resources they need.
On the other hand, it also increases the effects of ``noisy neighbors'', i.e., breaking the isolation of function containers~\cite{book_cloud_service_benchmarking}.
An approach to restrict CPU cores or even cycles is also feasible using Docker~\cite{Symeonides2020-hg,paper_hasenburg_mockfog2}, but would reverse that trade-off.

As discussed in C1 (Section~\ref{subsec:batch}), the auction-based approach suffers from the problem that requests will not arrive in batches unless the system is completely overloaded.
This means that AuctionWhisk will get individual requests and then has to decide whether to execute them or not.
We have implemented two different approaches for this: In the first, the Controller simply creates micro-batches by queuing requests in a short time window and then running an auction round over them.
This is as close as possible to the vanilla auction approach which we described in Section~\ref{sec:approach}.
The downside of this is that every request on average gets half the time window length as additional latency which will only be acceptable if the execution duration of requested functions is much greater than the time window length.

The second approach which we have implemented (and which we will later use in the experiments) avoids this extra latency and for this slightly deviates from the auction approach.
It introduces a new Decision Manager component within the Controller which uses a moving time window (default length one second) and tracks standard statistical aggregates such as average, median, min, max, etc. on the bids encountered in the time window (i.e., by default it knows the aggregates about all requests from the last second).
Users can then define a custom function that calculates a threshold value based on these values -- the default is the average.
If there is enough capacity (as described in the paragraph above), requests whose bid exceeds that threshold will be accepted, all others will be offloaded.
This is a bit different from the ``pure'' auction approach described in Section~\ref{sec:approach} but has the same or a very similar effect:
When the average bid of requested functions stays constant, the threshold will adapt itself based on load patterns until the node executes only the highest paying functions and is fully utilized.
When the average bid of requested functions increases, the threshold quickly adapts to the higher earning potential -- it lags slightly behind depending on the window length but achieves at least the same earnings as in the constant scenario.
When the average bid of requested functions decreases, the threshold also adapts quickly but may still reject a few requests near the threshold depending on the window length and how fast the average bid is decreasing.
Overall, we believe that this is a good solution which can easily be implemented in practice but, of course, alternative implementations that actually predict future requests and thus achieve higher earnings are possible.

\subsection{Experiment Setup}
\label{subsec:setup}

We also evaluated our approach through three experiments with AuctionWhisk.
All three experiments used the same setup:
The fog topology consists of one edge node, one intermediary fog node and one cloud node.
The hardware configuration of the nodes can be found in Table~\ref{tab:node_config}.
Please note that we used a relatively large edge node since OpenWhisk, the basis of AuctionWhisk, does not work well on smaller nodes~\cite{paper_pfandzelter_tinyfaas}.

\begin{table}
    \centering
    \begin{tabular}{ c | c | c | c |}
                    & Edge Node        & Intermediary Node & Cloud Node       \\\hline
        Description & CPX21            & CPX41             & CPX51            \\
        CPU         & 3 vCPU           & 8 vCPU            & 16 vCPU          \\
        Memory      & 4 GB RAM         & 16 GB RAM         & 32 GB RAM        \\
        Network     & 10 GBit          & 10 GBit           & 10 GBit          \\
        Storage     & NVMe SSD         & NVMe SSD          & NVMe SSD         \\
        OS          & Ubuntu 18.04 LTS & Ubuntu 18.04 LTS  & Ubuntu 18.04 LTS
    \end{tabular}
    \caption{System hardware configurations used for conducting the experiments.}
    \label{tab:node_config}
\end{table}

\begin{figure}
    \centering
    \includegraphics[width=0.6\columnwidth]{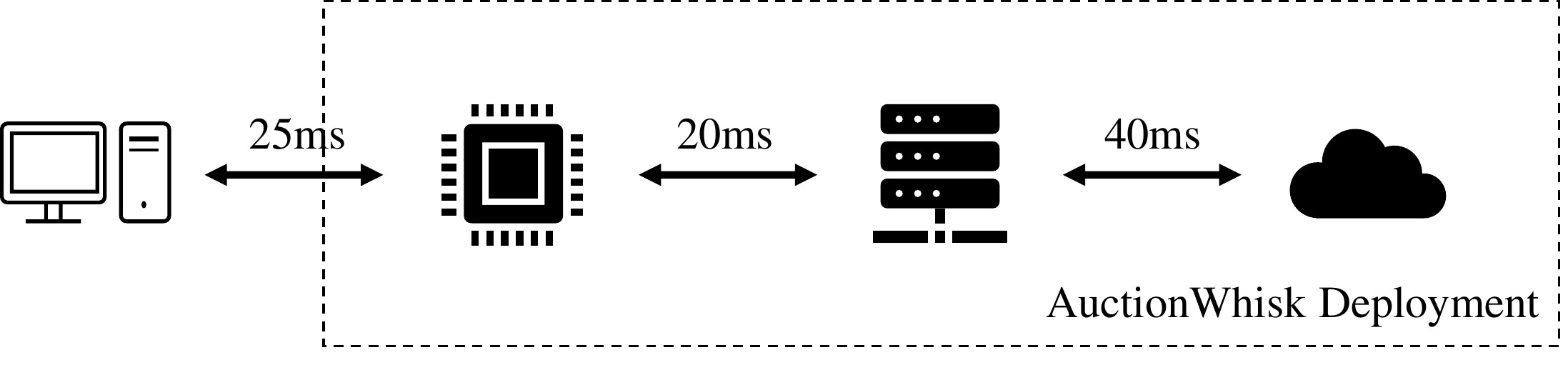}
    \caption{Experiment setup: average network latency from client devices (on the left) via edge and intermediary fog nodes to the cloud.}
    \label{fig:exp_ini_setup}
\end{figure}

Since we did not have access to a real fog infrastructure, we injected artificial latency based on netEm\footnote{man7.org/linux/man-pages/man8/tc-netem.8.html} similar to the approach used in MockFog~\cite{Hasenburg2019-er,paper_hasenburg_mockfog2} -- see Figure~\ref{fig:exp_ini_setup} for an overview.
Between the edge node and the intermediary node we set an average latency of 20 ms, the latency between the intermediary node and cloud node, we set to 40 ms, while our benchmarking client had a latency of 25 ms to the edge node (these numbers include both real and artificial delays).
On the edge node, we installed Lean AuctionWhisk, the intermediary and cloud node ran standard AuctionWhisk.

As discussed above, AuctionWhisk uses two parameters to track available compute capacity: A moving time window and the number of requests which can be accepted in said time window.
For the experiments, we set the time window length to 1000ms and limited the number of requests for that time window to 6, 12, and 1000 respectively (edge/intermediary/cloud).
We also increased the corresponding vanilla OpenWhisk parameters: For \texttt{invocationsPerMinute} we used 20,000 instead of 60 and for \texttt{concurrentInvocations} we used 240 instead of 30 to ensure that our experiments ran without failure.
Without changing these parameters, only 60 invocations per minute and 30 concurrent invocations would have been allowed.
These parameters were determined based on a number of initial experiments since standard OpenWhisk often has trouble coping with large numbers of parallel requests~\cite{paper_pfandzelter_tinyfaas} and we wanted to avoid frequent error responses in our experiments.

As experiment workload, we used a Node.js 10 implementation of the Sieve of Eratosthenes algorithm~\cite{sieve}, called with the argument 100, which was then automatically distributed over all nodes with a our deployment tool.
For creating load, we used Apache JMeter\footnote{jmeter.apache.org} and configured it to have several thread groups with several threads each which invoke the deployed functions for a period of 180 seconds.
As the concrete thread/thread group configuration varied between the experiments, we point out the exact numbers in the respective sections.

\subsection{Experiment 1: Request Latency of AuctionWhisk vs. a Cloud-Only OpenWhisk}
As the overall goal of the AuctionWhisk approach is to bring the benefits of edge and fog computing into the FaaS world, we ran a baseline experiment in which we compare request latency of AuctionWhisk to a cloud-only deployment of OpenWhisk.

We used the standard setting from Section~\ref{subsec:setup} and set our JMeter configuration to four thread groups.
For each thread group, we started with two threads and increased that number in steps of two up to 20 threads per thread group (80 threads in total).
Each thread issued a request every three seconds.
We deployed one function per thread group and assigned it a random bid (uniformly distributed over the interval [50;150]).
As a warm-up phase, we overloaded the target system for a short time to warm-up containers and to avoid cold starts during the actual experiment~\cite{mohan2019agile,paper_bermbach_faas_coldstarts}.
All experiments were repeated five times.

\begin{figure}
    \centering
    \begin{subfigure}{.6\columnwidth}
        \centering
        \includegraphics[width=\columnwidth]{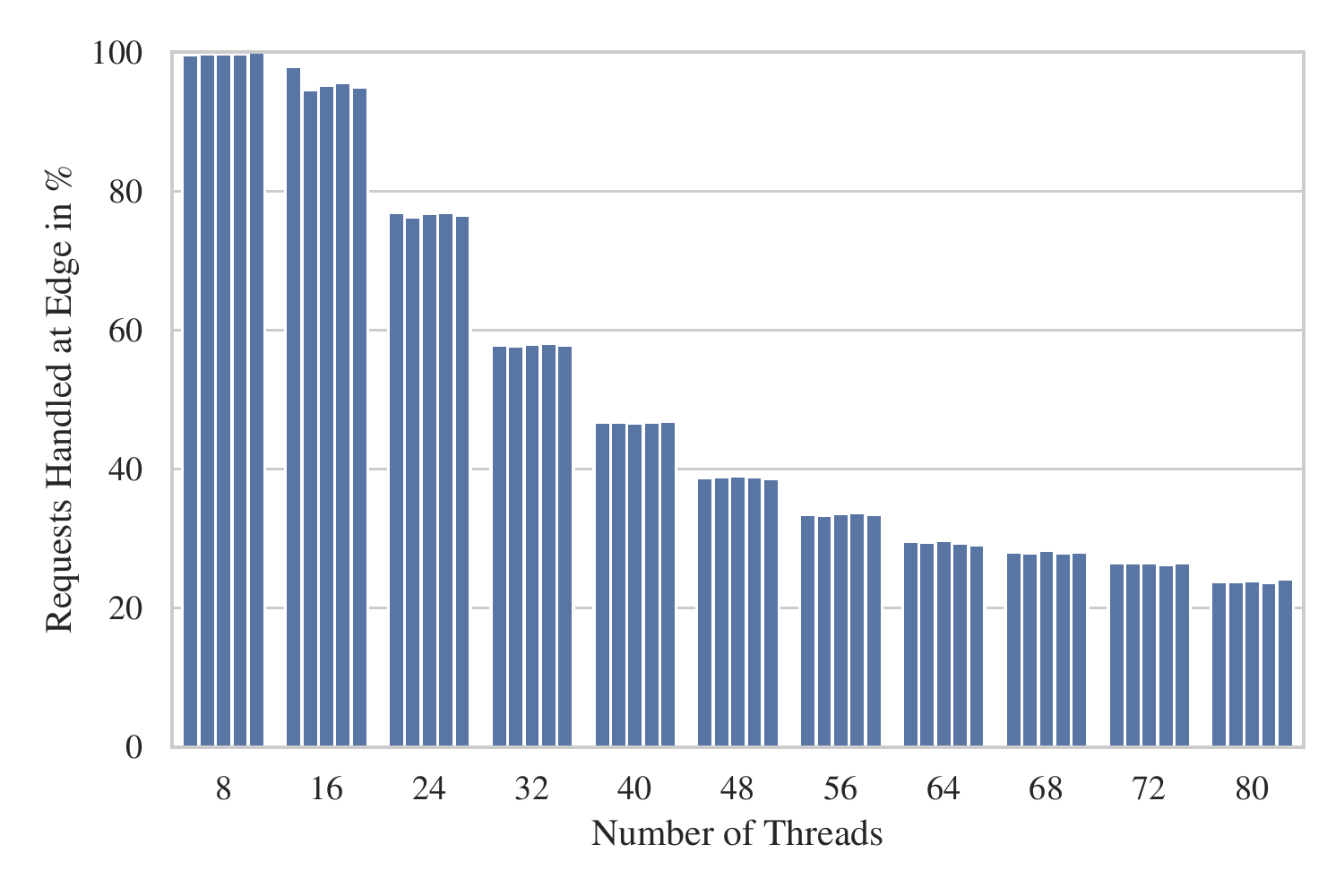}
        \caption{Impact of of increasing load on the percentage of requests that the edge can handle}
        \label{fig:exp_latency_sum_1}
    \end{subfigure}
    \hfill
    \begin{subfigure}{.6\columnwidth}
        \centering
        \includegraphics[width=\columnwidth]{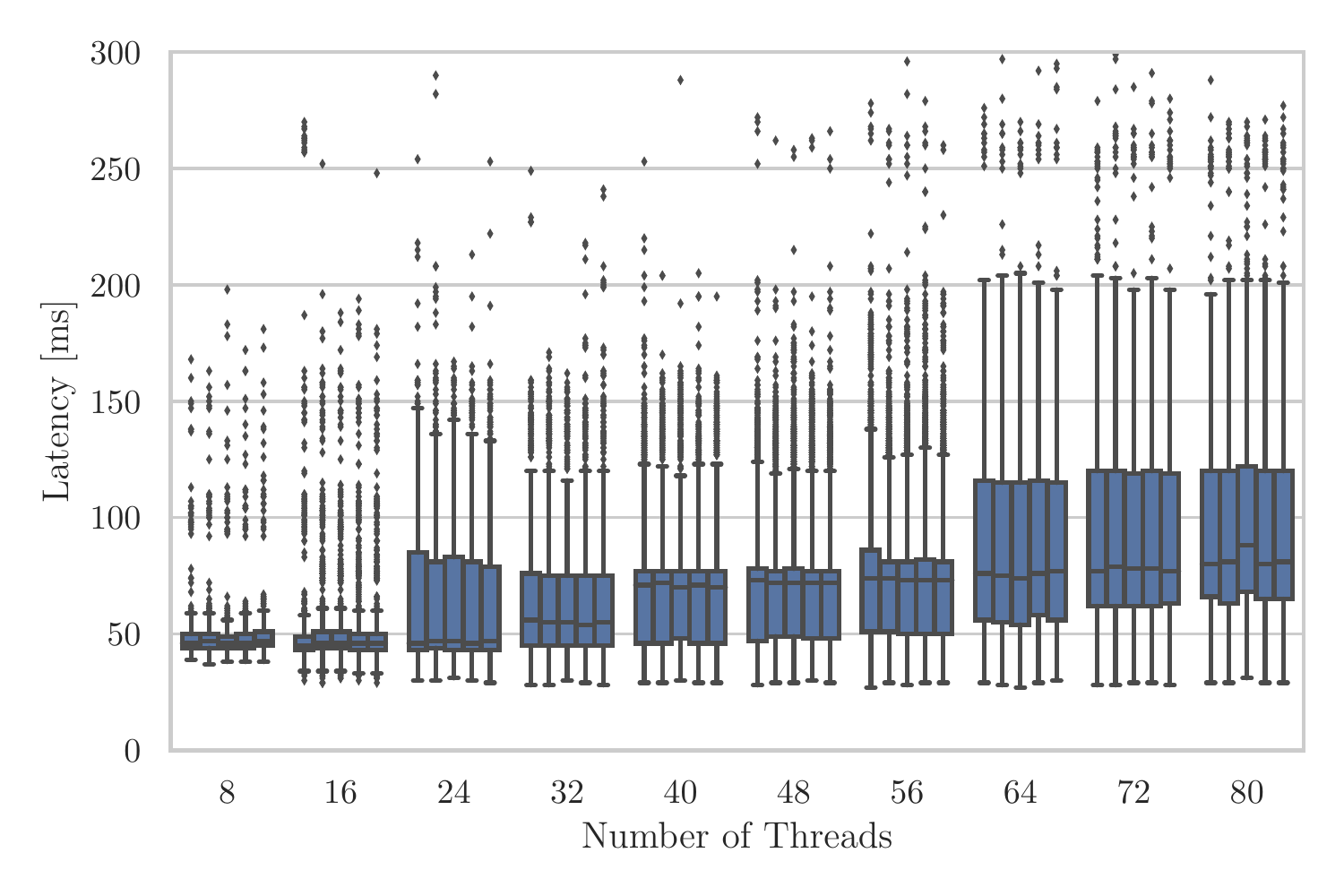}
        \caption{Impact on request latency as the load for AuctionWhisk is increased}
        \label{fig:exp_latency_sum_2}
    \end{subfigure}
    \caption{Experiment 1: Increasing the load for AuctionWhisk gradually increases request latency as the percentage of requests handled at the edge decreases and more and more requests are offloaded to the cloud. At approximately 16 threads, the edge node is saturated and starts offloading; at approximately 56 threads, the intermediary node starts offloading to the cloud node.}
    \label{fig:exp_latency_sum}
\end{figure}

\begin{figure}
    \centering
    \includegraphics[width=0.8\columnwidth]{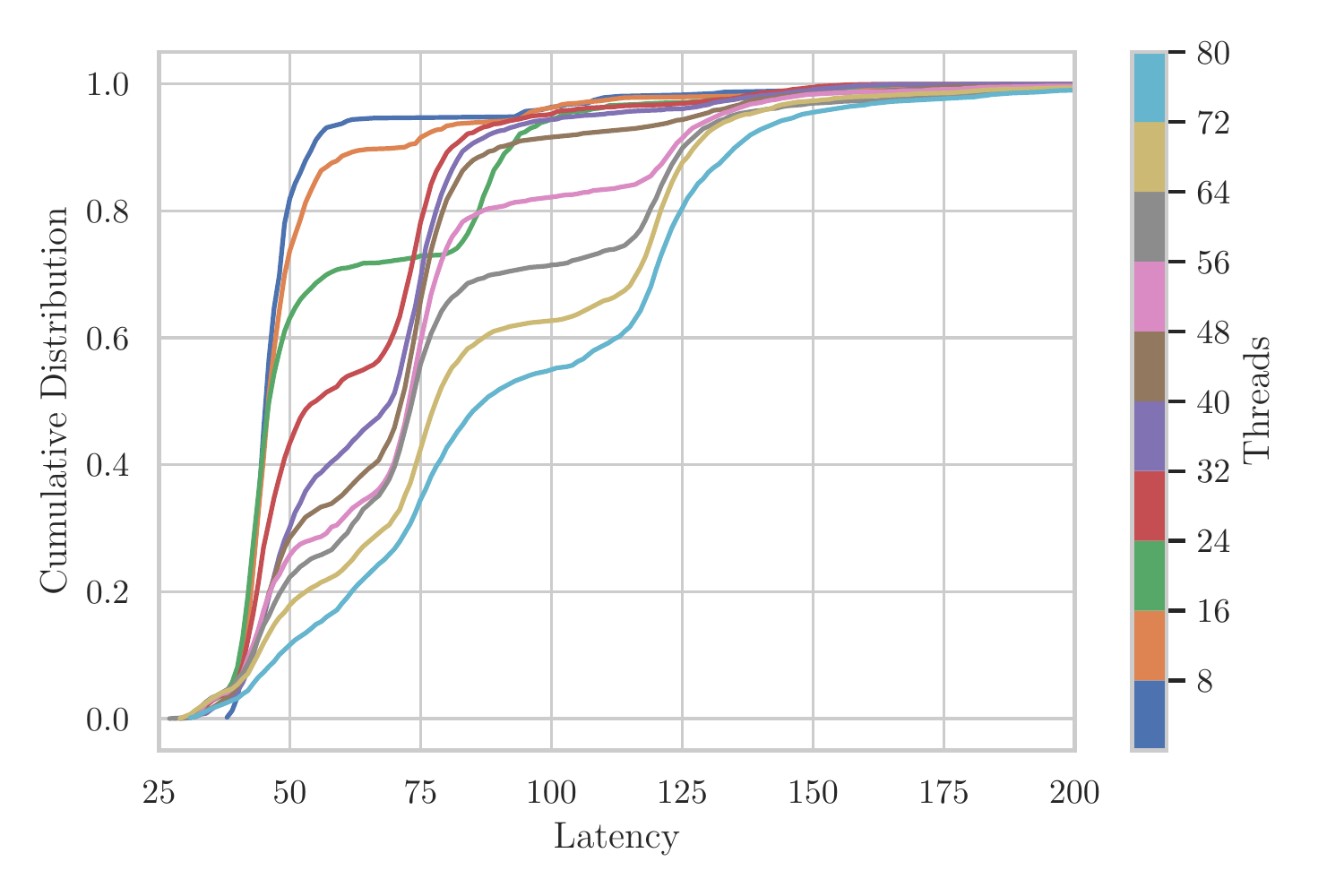}
    \caption{Experiment 1: CDF of AuctionWhisk runs at different load levels. The steep incline after ``plateaus'' indicates that any additional load is offloaded towards the cloud.}
    \label{fig:exp_auction_cdf}
\end{figure}

Figure~\ref{fig:exp_latency_sum} shows the results of all runs of experiment 1.
With an increasing number of threads, as more and more requests are offloaded towards the cloud (Figure~\ref{fig:exp_latency_sum_1}), the median latency increases (Figure~\ref{fig:exp_latency_sum_2}).
The first configuration with eight concurrent threads yields a latency of approximately 45ms, while all requests are handled at the edge.
The configuration with 16 threads shows that the median latency is only slightly higher.
However, the plot indicates that the edge slows starts offloading requests towards the intermediary node.
At 24 threads, the edge starts offloading a larger percentage of requests to the intermediary node, which leads to a higher 75th percentile and a higher whisker.
Running 40 threads, one can see that the median increases considerably.
This is due to the fact that at this point the edge offloads more than 50\% of all requests.
Further, we observe a second change in the 75th percentile once there are 64 threads running: At this time, the intermediary starts offloading a large percentage of requests to the cloud.
The outliers from the configuration with 56 threads indicate that single requests get offloaded at this point.

The CDF plot in Figure~\ref{fig:exp_auction_cdf} also confirms these observation:
The line for the experiment with 8 threads shows a high slope and then plateaus close to value of 1 -- this indicates that almost all requests are processed on the edge.
For all other configurations, we can see two ``plateaus'' (with a positive but low slope) which show the load levels which can still be sustained on one tier before offloading towards the cloud\footnote{If we had used $n$ tiers instead of three tiers in our experiment, we would have expected $(n-1)$ ``plateaus''.}.

\begin{figure}
    \centering
    \includegraphics[width=0.8\columnwidth]{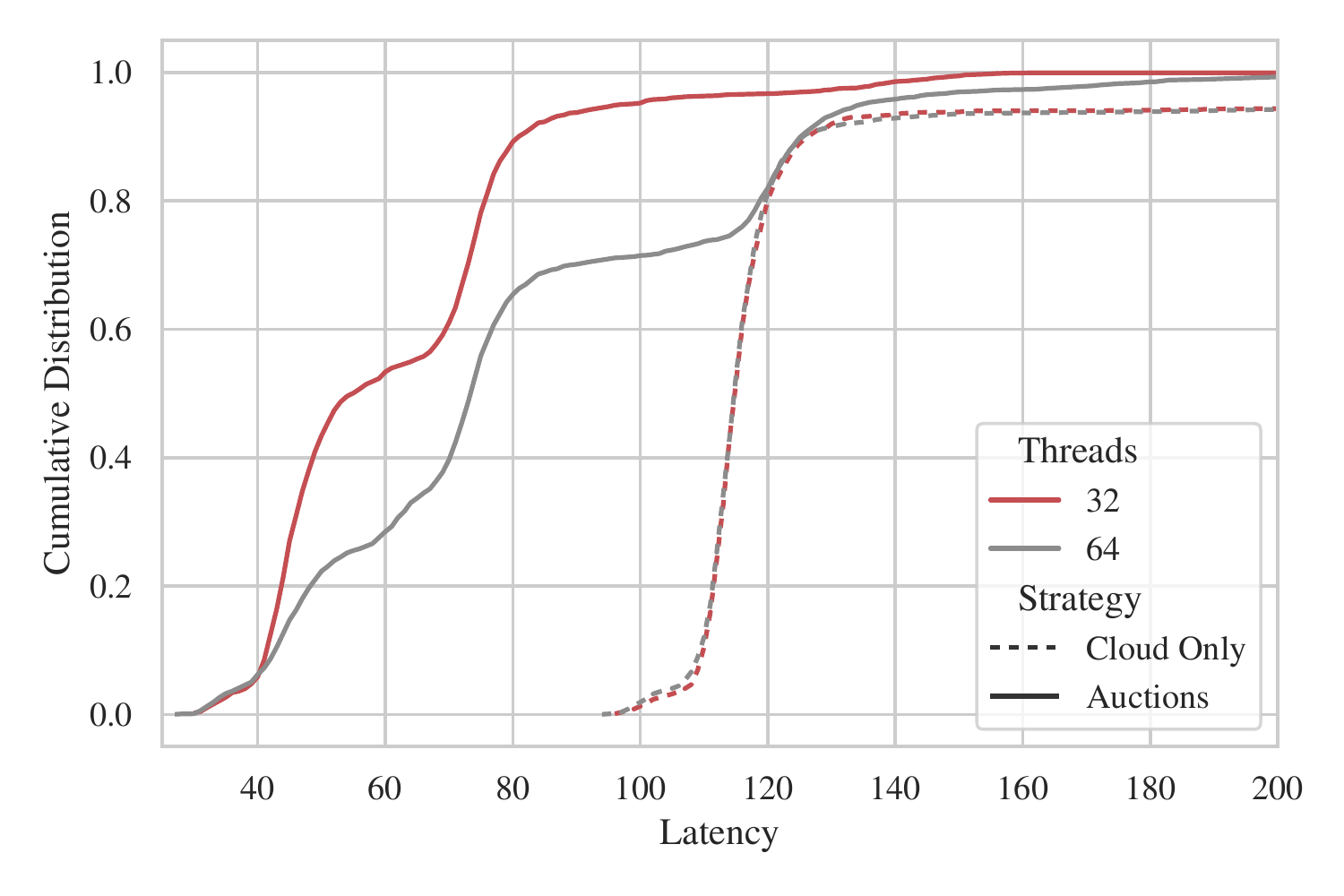}
    \caption{Experiment 1: CDF of AuctionWhisk and a cloud-only deployment of OpenWhisk for two different load levels. AuctionWhisk is at least as fast as OpenWhisk.}
    \label{fig:exp_latency_cloud_only_sum}
\end{figure}

For comparison, we also ran a cloud-only deployment with standard OpenWhisk in which we used a machine large enough to sustain the load level resulting from 64 threads.
To assert a fair comparison (see Figure~\ref{fig:exp_ini_setup}), we artificially delayed requests by 85ms (=25ms + 20ms + 40ms) and ran the exact same workload as in the AuctionWhisk experiments with 32 and 64 threads.
As we can see in Figure~\ref{fig:exp_latency_cloud_only_sum}, AuctionWhisk serves requests at least as fast as the cloud-only deployment of OpenWhisk\footnote{The CDF does not reach the level of 1 in our chart as OpenWhisk often suffers from latency spikes and failed requests~\cite{paper_pfandzelter_tinyfaas}.}.
While this is as expected, it shows that the AuctionWhisk approach allows OpenWhisk to efficiently leverage resources at or near the edge.

\begin{figure}
    \centering
    \begin{subfigure}{0.6\columnwidth}
        \centering
        \includegraphics[width=\columnwidth]{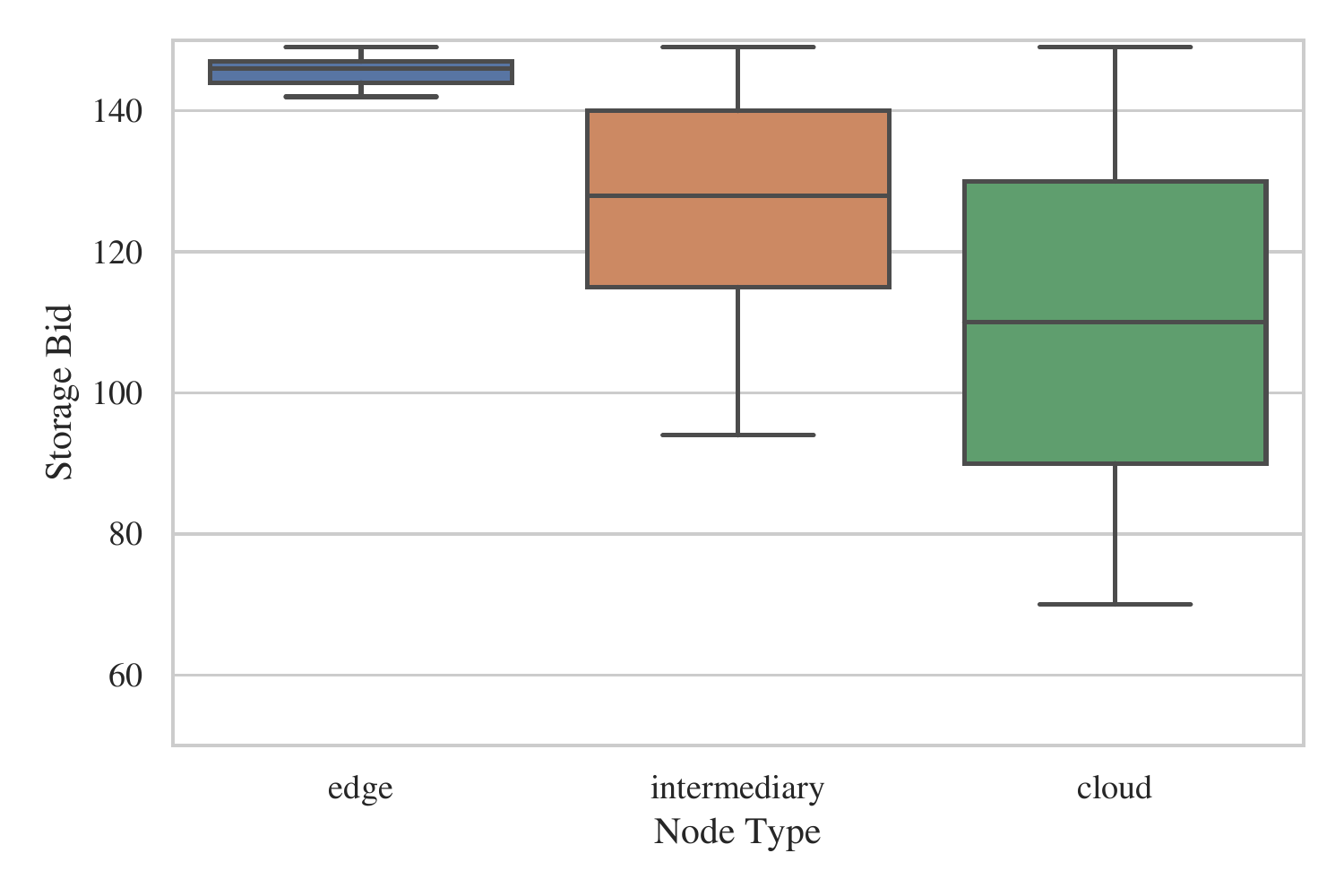}
        \caption{Clients pay the same bid on all node types. [calculated]}
        \label{fig:storage_1}
    \end{subfigure}
    \hfill
    \begin{subfigure}{0.6\columnwidth}
        \centering
        \includegraphics[width=\columnwidth]{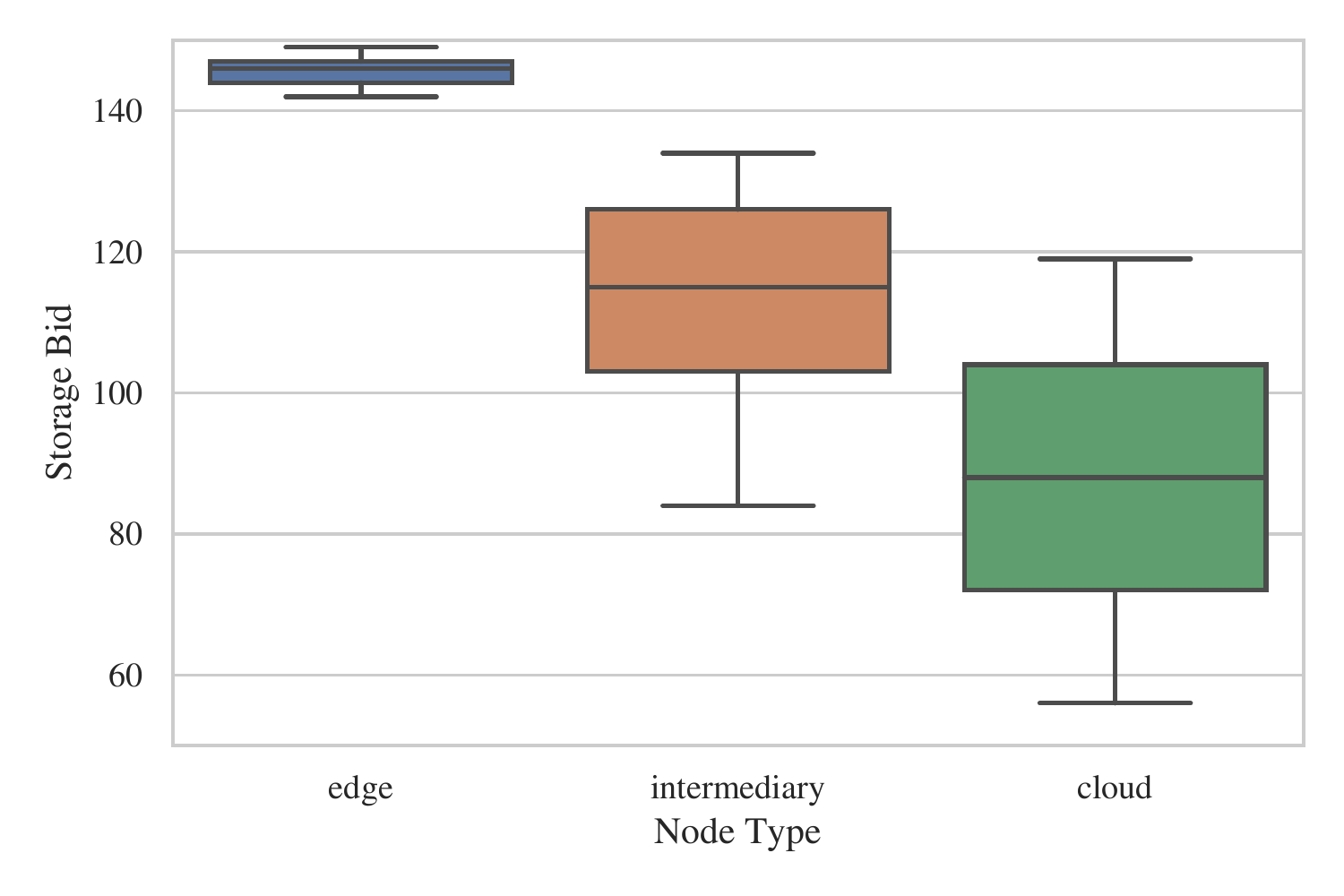}
        \caption{Clients use a discounted bid in the cloud (80\%) and on intermediary nodes (90\%). [measured]}
        \label{fig:storage_2}
    \end{subfigure}
    \caption{Experiment 2: Distribution of successful bids on different node types. Nodes with limited capacity at or near the edge only store the executables with the highest bids while the cloud stores all executables.}
    \label{fig:storage}
\end{figure}

\subsection{Experiment 2: Effect of Storage Prices}
The setup of this experiment is very similar to Simulation Experiment 2 from Section~\ref{sec:eval_sim}: We analyze the effect of storage bids on function placement and design the experiment in a way that asserts that the processing bids do not affect the outcome.
In the experiment, we deployed the exact same function 1000 times with a random bid from the interval [50;150].
To study storage bid effects in an isolated way, we artificially limited the storage capacity of the edge and intermediary node to assert that they could store only 100 and 500 functions respectively.
In practice, developers can be expected to use higher bids for the edge or the intermediary (to increase the chance of their function executing there).
For this experiment, we implemented this as a discount: functions with a bid $b$ will pay $b$ to edge nodes, $0.9\cdot b$ to intermediary nodes, and $0.8\cdot b$ to the cloud.
We used detailed logging so that we could also calculate the node earnings without the discounted bids and repeated the experiment three times.
Through previous test runs, we had already asserted that requests are indeed offloaded when an executable is not found locally and therefore decided not to run a workload against this deployment.

Figure~\ref{fig:storage} shows the results of one experiment run with (Figure~\ref{fig:storage_2}) and without (Figure~\ref{fig:storage_1}) discounted bids.
As expected, we can see that functions stored at the edge have an average storage bid close to the upper bound.
This is due to the restricted storage space, which forces the node to remove all functions with lower storage bids.
In contrast to this, the cloud node stores all functions, from high paying to low paying -- this can be clearly seen in Figure~\ref{fig:storage_1} where the upper whiskers for all node types have the same value.
In general, both figures demonstrate the desired effect while the discounted bids (which could also be implemented as a surcharge on the edge) further aggravate the effect.
The exact same effect can be seen in the other two test runs which, however, had slightly different values due to the randomization of bids.

\subsection{Experiment 3: Effect of Execution Prices}
The setup of this experiment is very similar to Simulation Experiment 1 from Section~\ref{sec:eval_sim}: We analyze the effect of processing bids on function placement and design the experiment in a way that asserts that the storage bids do not affect the outcome.
We deployed the same function 100 times, each with a random processing bid in the interval [50;150].
For the workload, we used eight parallel thread groups (with eight threads per thread group) which each targeted one randomly selected function, i.e., we invoked eight different functions which each had their respective processing bid.
We repeated the experiment three times.

\begin{figure}
    \centering
    \begin{subfigure}{.5\textwidth}
        \centering
        \includegraphics[width=\columnwidth]{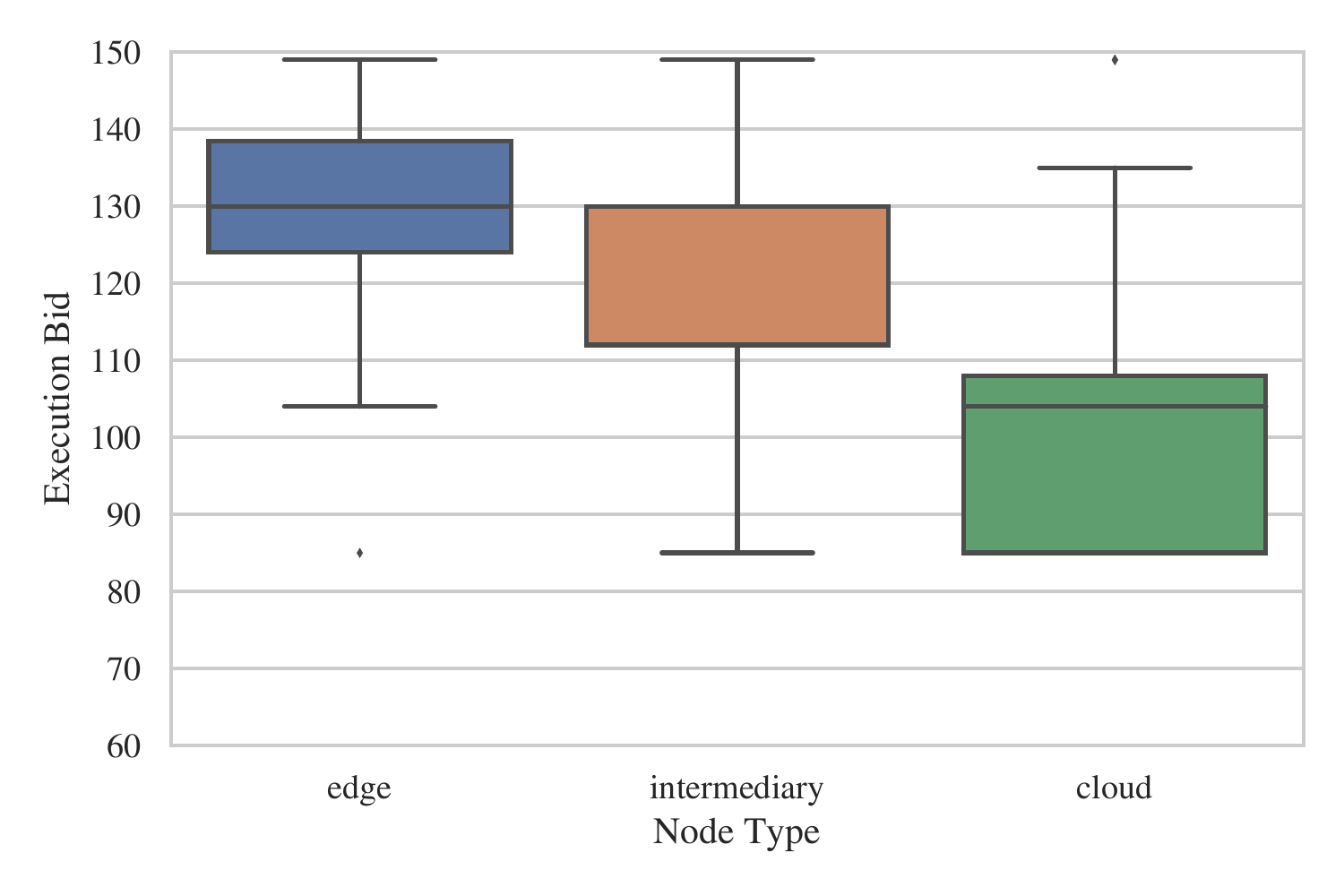}
        \caption{Clients pay the same bid on all node types. [measured]}
        \label{fig:exec_1}
    \end{subfigure}
    \hfill
    \begin{subfigure}{0.6\columnwidth}
        \centering
        \includegraphics[width=\columnwidth]{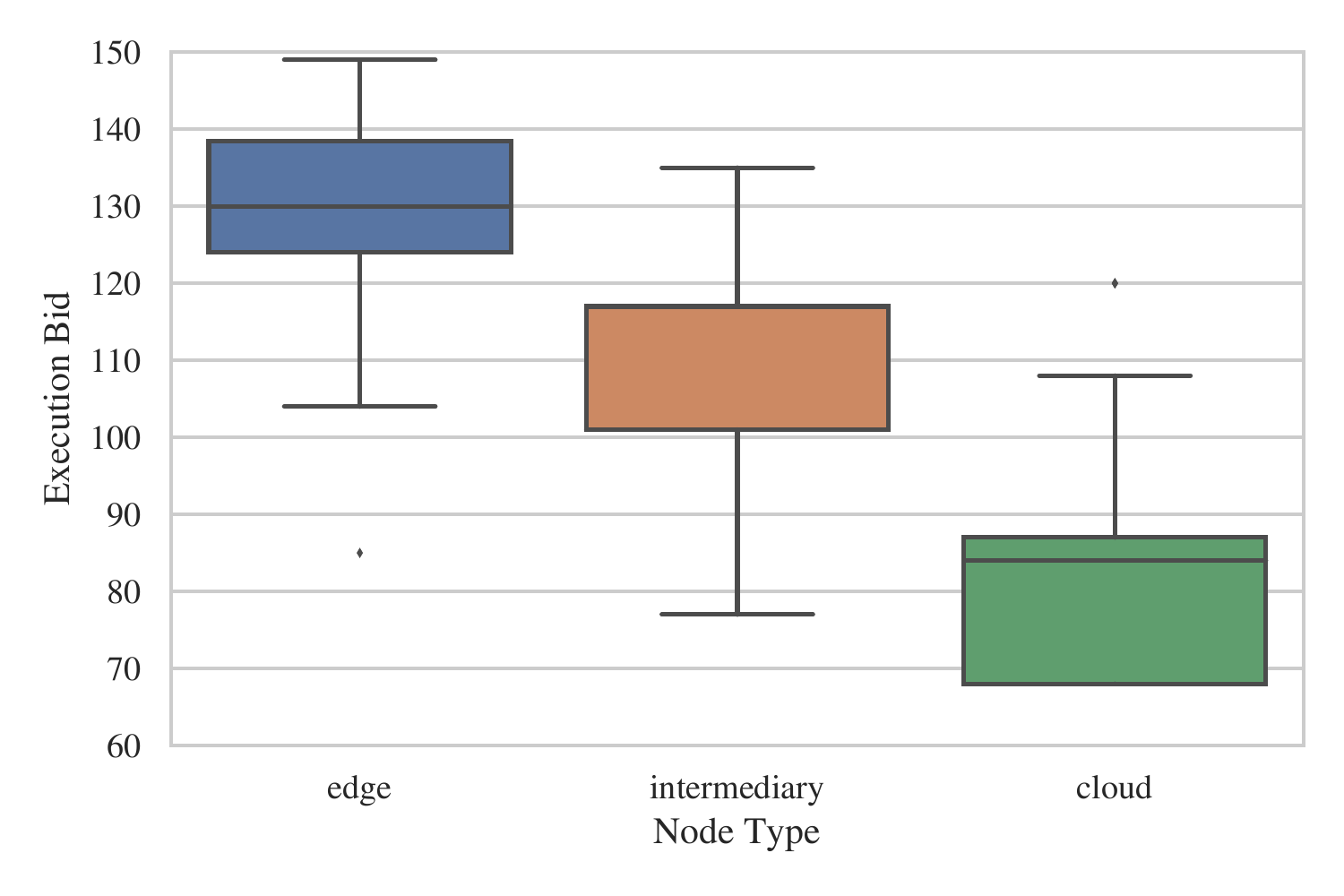}
        \caption{Clients use a discounted bid in the cloud (80\%) and on intermediary nodes (90\%). [calculated]}
        \label{fig:exec_2}
    \end{subfigure}
    \caption{Experiment 3: Distribution of prices paid per execution across different node types. The higher the processing bid of the requested function is, the more likely is the request to be executed at or near the edge.}
    \label{fig:exec}
\end{figure}

The results of one experiment run are shown in Figure~\ref{fig:exec}:
Similar to experiment 2, we not only show the results of the base setup as described above (Figure~\ref{fig:exec_1}) but also calculated the outcome for developer discounts on intermediary nodes (90\% of the edge price) and in the cloud (80\% of the edge price) -- see Figure~\ref{fig:exec_2}.
As expected, we can see that the distribution of prices paid for execution strongly increases towards the edge as the nodes with limited resources are able to cherry-pick the highest paying requests only.
With the discount option, this effect is even more apparent.
The other two experiment runs confirm these results but have slightly different values due to the randomization when selecting the 8 out of 100 target functions.

Overall, our experiments with AuctionWhisk show the desired effect of an efficient resource allocation where the location of a function execution is determined by the price the application developer is willing to pay.

\section{Discussion}\label{sec:discussion}
With our simulation models and empirical measurements on our prototype AuctionWhisk, we have shown that it is feasible to place functions in fog-based FaaS platforms using decentralized auction-inspired mechanisms.
In the following, we will discuss limitations of our approach and derive avenues for future work.

\textbf{Bidding of application developers.}
In our experiments, we assume simple bidding strategies from application developers, yet more advanced strategies, such as developers providing high bids for scarce resources that are urgently needed while offering low bids for, e.g., processing and storage in the cloud, are also possible.

Although this is not the focus of our research, our auction-inspired approach also invites a game-theoretic perspective, e.g. on combining ostorage and processing bids such as to use a high storage bid with a bloated executable to effectively get a single tenant edge node combined with a very low processing bid to reduce cost.

Furthermore, as bids are provided at deploy time, they are relatively static.
This means that developers cannot easily take advantage of periods of low demand (e.g., at night) with temporarily lower bids and reduced overall costs.
Also, actual prices may vary a lot over time depending on the actions of other developers as well as the request rates created by end users.
The result is that earnings at the edge are maximized, with the goal of making the high CapEx of installing edge infrastructure worthwhile, while tenants can specify how much they are willing to pay.

As an additional factor, the impact of providing information about the location of function execution, system load, or bids of other tenants on optimal bidding strategies may be an interesting subject for further research.

\textbf{Auction mechanism.}
In our simulation model, nodes are rather simplistic in that they do not look into the future and do not couple storage and processing bid -- they simply accept all executables that fit in and serve whatever is received in the order of the bids.
While this is in line with our chosen auction model, much higher revenue could be achieved through decisions that follow pricing trends.
For instance, just considering the processing bid together with past invocation numbers and coupled with the storage bid could result in considerably higher profits.
As a more concrete example, it may make sense to delegate a processing request even though the node is not saturated when a periodic request with a higher processing cost is expected to arrive shortly.

Likewise, our AuctionWhisk prototype considers both bids separately through the heuristics described in Section~\ref{sec:prototype}.
We decided not to use the micro-batching approach for the execution auction in our experiments as it increases the overall latency significantly.
Nevertheless, micro-batching would further increase the earnings for the respective node.
For a deployment in practice, we suggest to use micro-batching when the latency to the next node on the path to the cloud is high and our proposed heuristic in all other cases.

\textbf{Overall resource allocation.}
Combining both sides of the auction, significant improvements are possible when application developers try to minimize cost while nodes try to maximize earnings.
However, even then the overall solution will not yield the globally optimal result that could be achieved through an allocation optimization across all nodes.
Yet, in exchange for optimizing only locally, nodes do not need to communicate for their decision, leading to faster decisions and thus lower end-to-end latency, more overall scalability, and potentially also an increased resiliency against connection losses.
In some scenarios, it may also be beneficial not to offload a request but rather to queue it on the edge briefly (thus possibly reducing end-to-end latency while keeping the earnings for the edge node).

\textbf{No guarantees for application developers.}
Overall, our solution takes a rather particular approach in that application developers get no guarantees where their function will be executed -- the placement of functions only depends on the bids at a specific point in time.
There is, hence, no way to provide bounds on response time and other quality dimensions, i.e., the infrastructure provider does not provide any guarantees beyond \emph{eventually} executing every function \emph{somewhere}.
In a way, this is comparable to airline seat assignment schemes in which economy class passengers can bid for business class upgrades.
While this is undesirable from an application developer perspective, our proposed approach has the benefit that it requires no centralized coordination at all, i.e., it scales well and is resilient, and that it is likely to maximize earnings for the infrastructure provider.
Also, the lack of guarantees is in line with the state-of-practice in cloud services which provide minimal to no guarantees as part of their SLAs~\cite{bradshaw2011contracts,book_cloud_service_benchmarking}.

Still, if our proposed approach were to be used in practice, we would expect infrastructure providers to let customers reserve subsets of their resources at a fixed price -- comparable to 5G slicing~\cite{foukas2017network} -- and use the proposed auctioning scheme only for a subset of their resources.
This would lead to the interesting situation that application developers that actually need quality guarantees can pay for it (comparable to directly booking a business class seat) while others can still bid on the remaining resources (comparable to bidding on a business class upgrade).

Finally, developers can currently not get a guarantee that function execution happens only in a certain area to, e.g., satisfy regulatory or privacy requirements.
While they could decide to send their executable to only a limited subset of nodes, requests that are outbid on the edge will still be propagated to the cloud before the cloud node detects that it does not have an executable for the request.
With personal data of end users governed by the GDPR, this might be illegal depending on the location of the cloud node.
One way to address this would be to let developers not only specify bids but also constraint individual offloading paths of functions.
This way, a request could be offloaded only to desired nodes (possibly nearby edge nodes).

\textbf{Implementation challenges.}
There are of course a number of technical challenges in implementing a fully integrated platform that we did neither address with our concept nor our prototype implementation.
For instance, we did not consider node failures or network partitioning:
The former case could be addressed by using a reliable messaging middleware for communication between nodes, which would ensure that all requests are eventually delivered to a node for execution.
In the case of network partitioning, offloaded requests will be lost in our current prototype.
We do, however, not see this as a major limitation as, in such a scenario, some requests \emph{will have to} be dropped.

There is also the aspect of co-managing data placement~\cite{paper_hasenburg_towards_fbase,techreport_hasenburg_2019,paper_zhang_cloud_is_not_enough_GDP,paper_confais_ipfs4fog} and function placement~\cite{paper_pfandzelter_streams_functions,Hasenburg2018-nd,paper_baresi_serverless4fog,pfandzelter-zero2fog-wiley}, which we think is a promising avenue for future research~\cite{paper_hellerstein_serverless,sreekanti2020cloudburst}.

Finally, we did not consider undeployment of functions:
Depending on storage demand, this may remove the executable from an otherwise fully utilized node.
The node, however, then no longer knows of rejected or evicted executables, thus potentially reducing future earnings in this node.
This problem could be addressed by having a repository where developers submit their executables and bids as a single source of truth that nodes could then query to gain access to executables even after evictions.

\textbf{Summary.}
While there are some limitations in our approach, we believe its benefits outweigh the disadvantages.
Most of the limitations could also be addressed through compensation mechanisms or small extensions.
Furthermore, the overall AuctionWhisk approach opens up a range of interesting game-theoretic research questions, which are beyond the scope of this paper but worthy of further research.

\section{Related Work}\label{sec:relwork}
The service placement problem is inherent to fog computing, and deploying distributed applications over a distributed, heterogeneous infrastructure under constraints such as bandwidth usage, latency limits, or regulatory requirements is a combinatorial problem.
Consequently, this challenge has been a key research topic in the last few years.
On the one hand, there are static solutions using upfront best practices~\cite{paper_pfandzelter_streams_functions,Gusev2019-ch,Karagiannis2020-kx,Santos2020-qx}, simulation~\cite{Hasenburg2018-fn,Hasenburg2018-nd,Gupta2017-jx,spe.2787,Brambilla2014-kl,Sotiriadis2014-uz,Zeng2017-lq,Qayyum2018-ml,Fernandez-Cerero2020-oa,Sonmez2018-sq,Giang2015-ws}, testbed evaluation~\cite{Hasenburg2019-er,Mayer2017-dt,Coutinho2018-wa,Eisele2017-yx,Mayer2017-dt,Coutinho2018-wa,Banzai2010-lr,De_Oliveira2014-bq,Balasubramanian2014-tj,luckow2021exploring}, a combination of those~\cite{pfandzelter-zero2fog-wiley,Roy2011-uy}, or formalized assignment problems~\cite{de2019multi,Brogi2017-fk,cardellini2017optimal,Oh2020-xj,8960404,8897679,spe.2951,Khare2019-ef,Xu2019-rq,9358007,Janssen2018fogflink}.
On the other hand, dynamic approaches using centralized schedulers~\cite{basic2019fuzzy,Tong2016-ke,9132684,8975987,9305277,9339982,8955944,8869806,spe.2896,Naas2017-ln,Wobker2018-ud,spe.2966} or decentralized algorithms~\cite{9361310,9373980,Saurez2016-oo,becker2021los} have also been proposed.
It is no surprise that auction-based approaches, which have seen broader interest in computer science~\cite{Huang08:ABR,Gao11:MAP,Zhao14:HCT,Yang16:IMC,Kantarci14:TSP,Zhang14:DRP,Chandrashekar07:ABM,Jayaweera11:ACC}, have also been proposed for fog computing, especially in those multi-provider, multi-tenant scenarios where both infrastructure providers and application service providers want to find a cost-optimal solution to service placement~\cite{fawcett2016combinatorial,MAHMUD2020177,ZHOU2019151,8896972,8657771,7536541,spe.2839,9205640,8254400,9042359,8489932,8306104,8410767,9163371,8030317,8449792,8648013,10.1145/3390557.3394130,7858574,peng2020multiattribute,8904239,9342683,jiao2019auction,luong2020machine,10.1007/978-3-030-44041-1_25,7972940}.
For example, Fawcett et al.~\cite{fawcett2016combinatorial} present an auction-based resource allocation platform for the fog that has service providers interested in using the fog infrastructure bid for resources provided by independent infrastructure providers.
They employ combinatorial bids for combinations of CPU, storage, and memory for a specified time slot and assume services to be available in the form of containers.
A central orchestrator acts as an auctioneer towards service providers and collects their bids, then performs a provisioning step where it provisions the services of the winning bids on the available devices.

Nevertheless, existing research has not taken into account the unique challenges and opportunities of fog-based FaaS platforms, although it has been shown that the FaaS paradigm is a good fit for fog applications~\cite{paper_pfandzelter_streams_functions,paper_pfandzelter_tinyfaas,wang2021lass,paper_bermbach_fog_computing}.
Notably, the finely-grained execution units of serverless functions can lead to high load on central scheduling components, as Rausch et al.~\cite{paper_rausch_towards_serverless4edge} have shown in comparison to their earlier work~\cite{paper_nastic_serverless_at_edge,nastic2018towards,RAUSCH2021259}.
The authors claim that there is no technical solution that is able to handle such scheduling of function execution across edge, intermediaries, and the cloud.
On the other hand, precisely this flexible scheduling of function executions can maximize resource allocation and lead to better auction conditions for both providers as well as consumers.
Using long-running services that are migrated with user movement (e.g., as suggested by~\cite{paper_machen_live_migration_mec,paper_wang_mec_service_migration}) blocks scarce resources on constrained and expensive edge and fog nodes and is thus less efficient than executing functions only when needed.

Nevertheless, some scheduling solutions using centralized monitoring or control units have been proposed:
For example, Aske and Zhao~\cite{aske2018supporting} propose a central monitoring solution that collects telemetry data from different fog FaaS platforms and schedules functions to the best platform given per-function constraints.
However, this leaves the majority of scheduling work with the platform providers as it abstracts from the actual platforms.
Cheng et al.~\cite{cheng2019fog} introduce function coordination based on ``data,'' ``usage,'' and ``system'' contexts to optimize data-intensive applications.
To some extent, this assumes scalable infrastructure as resources need to be available where data is produced, and a central coordinator needs to keep track of all data streams.
Furthermore, as this approach is tailored to data-intensive applications, it can probably not deal with event-driven applications.
Pelle et al.~\cite{pelle2020telemetry} as well as Palade et al.~\cite{palade2020swarm} use centralized controllers for function scheduling, albeit not on a per-execution basis.
Nevertheless, this requires a central, global view of the FaaS system, which is especially unfeasible in a multi-provider fog setup.

Decentralized approaches forego such a central component to increase scalability and availability.
Persson and Angelsmark~\cite{persson2017kappa} build their \emph{Kappa} platform on top of \emph{Calvin}~\cite{persson2015calvin}, as distributed IoT platform.
They deploy functions as applications on top of Calvin, yet it is unclear how function scheduling is achieved beyond static assignment of functions to compute nodes.
Baresi and Medonca~\cite{paper_baresi_serverless4fog} propose a serverless edge platform that comprises independent FaaS platforms distributed across the fog.
The platforms coordinate function offloading among each other if needed, based on current load and request QoS requirements.
Although their approach avoids a centralized scheduler that could introduce a bottleneck or single point of failure, it is unclear on what basis function offloading is decided.
Regional load balancers distribute requests within their region, yet the paper remains vague regarding the question whether multiple such load balancers exist or what their impact is on request latency, as additional network hops introduce overhead.
In their earlier work~\cite{baresi2017empowering}, clients themselves would need to specify whether to execute a function at a local edge node or in the cloud.
This is also a common approach in commercial solutions for edge function execution.
Most notably, AWS offers Greengrass\footnote{aws.amazon.com/greengrass} and Lambda@Edge\footnote{aws.amazon.com/lambda/edge}, while Microsoft has added Azure Stack Edge\footnote{azure.microsoft.com/en-us/products/azure-stack/edge} to their portfolio.
These solutions require application developers to address function execution environments at the edge or in the cloud in different ways, so that the burden of resource management is left to developers.
To this end, Das et al.~\cite{das2020performance} propose client-side models that predict whether an execution on the edge or in the cloud is more efficient to satisfy given constraints.
Cicconetti et al.~\cite{cicconetti2020uncoordinated} argue for uncoordinated access, where clients are given a list of possible function execution locations and make local decisions, for example through probing different options and adapting over time.
While this does not lead to a global optimum, the scheduling load on clients and infrastructure is low.
Two assumptions make this a worse deal for infrastructure providers than our auction-based approach: first, all possible function nodes must store the function code at all times as they do not know when a client changes their decision on execution location.
Second, an edge node cannot influence when it would like to be considered as an execution location for a function, for instance once it has enough resources available.
This prohibits a dynamic pricing policy.
The authors have also proposed using intermediary routers~\cite{cicconetti2020decentralized} as decentralized function schedulers, an approach where both clients and nodes have no control over function execution.

\section{Conclusion}\label{sec:concl}
Serverless FaaS is a promising paradigm for fog environments.
In practice, fog-based FaaS platforms have to schedule the execution of functions across multiple geo-distributed sites, especially when edge nodes are overloaded.
Existing approaches mostly argue for centralized placement decisions which, however, will not scale indefinitely~\cite{paper_rausch_towards_serverless4edge}.

In this paper, we extend a previous paper~\cite{paper_bermbach_auctions4function_placement} and follow an auction-inspired scheme in which application developers specify bids for storing executables and executing functions across the fog.
This way, overloaded fog nodes can make local decisions about function execution.
Therefore, also the function placement is distributed across nodes and no longer the scalability bottleneck it is with centralized scheduling approaches.
Furthermore, we showed through simulation that our approach asserts that all requests are served while maximizing revenue for overloaded nodes.
We also showed that such an approach can be implemented in real-world FaaS platforms by extending the open source system Apache OpenWhisk for our AuctionWhisk prototype and conducting three experiments with it.

\section*{Acknowledgments}
We thank Setareh Maghsudi for contributing an overview of various auction approaches to our original paper~\cite{paper_bermbach_auctions4function_placement}.

\bibliographystyle{plain}
\bibliography{bibliography}

\end{document}